\RequirePackage{graphicx}
\documentclass[prd, superscriptaddress, nofootinbib, floatfix]{revtex4-2}  
\usepackage{enumitem}% http://ctan.org/pkg/enumitem
\usepackage{mathtools}
\usepackage{tikz}
\usepackage{xcolor}
\usepackage[subfigure]{graphfig}
\usepackage[normalem]{ulem} %to strike the words
\usepackage{wrapfig}
\usepackage{amsmath}
\usepackage{amsfonts}
\usepackage{amssymb}
\usepackage{booktabs}
\usepackage{multirow}
\usepackage{slashed}
\usepackage{physics}
\usepackage{tabularx}
\usepackage{soul}
\usepackage{ulem}
\usepackage{hhline}
\usepackage{float}
\usepackage{enumitem}
\usepackage[colorlinks,pdfusetitle]{hyperref}
\hypersetup{colorlinks=true,allcolors=[rgb]{0,0.0,1.0}}
\usepackage{longtable}
\usepackage{relsize}
\usepackage{verbatim}

\usepackage{nccmath}
\usepackage{esvect}

\newcommand\befs{\begin{figure*}[ht!]}
\newcommand\eefs[1]{\label{fig:#1}\end{figure*}}
\newcommand\bef{\begin{figure}[ht!]}
\newcommand\eef[1]{\label{fig:#1}\end{figure}}
\newcommand\beq{\begin{equation}}
\newcommand\eeq[1]{\label{#1}\end{equation}}
\newcommand\beqa{\begin{eqnarray}}
\newcommand\eeqa[1]{\label{#1}\end{eqnarray}}
\newcommand\bet{\begin{table}}
\newcommand\eet[1]{\label{tb:#1}\end{table}}
\newcommand\bets{\begin{table*}}
\newcommand\eets[1]{\label{tb:#1}\end{table*}}

\def\be{\begin{equation}}
\def\ee{\end{equation}}
\newcommand{\bea}{\begin{eqnarray}}
\newcommand{\eea}{\end{eqnarray}}
\newcommand{\ba}{\begin{align}}
\newcommand{\ea}{\end{align}}

\newcommand{\Dl}{\Delta}
\newcommand{\wt}{\widetilde}

\usepackage{graphicx}
\graphicspath{{figures/}}

\begin{document}

\widetext

\title{Gluon unpolarized, polarized, and transversity GPDs from lattice QCD: Lorentz-covariant parametrization} 

\newcommand{\RBRC}{RIKEN-BNL Research Center, Brookhaven National Laboratory, Upton, NY 11973, USA}\affiliation{\RBRC} 
\newcommand{\BNL}{Physics Department, Brookhaven National Laboratory, Upton, NY 11973, USA}\affiliation{\BNL}
\newcommand{\NMSU}{Department of Physics, New Mexico State University, Las Cruces, NM 88003, USA}\affiliation{\NMSU}
\newcommand{\CAS}{CAS  Key  Laboratory  of  Theoretical  Physics,  Institute  of  Theoretical  Physics, Chinese  Academy  of  Sciences,  Beijing  100190,  China.}\affiliation{\CAS}

%%%%%%%%%%%%%%%%%%%%%%%%%%%
% AUTH BLOCK
%%%%%%%%%%%%%%%%%%%%%%%%%%%
\author{Jakob~Schoenleber}\affiliation{\RBRC}\affiliation{\BNL}
\author{Raza~Sabbir~Sufian}
\affiliation{\NMSU}\affiliation{\RBRC}\affiliation{\BNL}
\author{Taku~Izubuchi}\affiliation{\RBRC}\affiliation{\BNL}
\author{Yi-Bo Yang}\affiliation{\CAS}

\begin{abstract}
%%%%%
We identify the matrix elements necessary to determine the  leading-twist gluon generalized parton distributions (GPDs)  $H_g,~E_g,~\wt{H}_g,~\wt{E}_g,~H^T_g,~E^T_g,  \wt{H}^T_g ,~\wt{E}^T_g$ in lattice QCD calculations. We present a method to achieve a Lorentz-covariant parameterization of the matrix elements in terms of  a linearly independent basis of tensor structures. This parameterization is crucial for projecting lattice QCD matrix elements onto light cone distributions. For the first time, we determine the corresponding components that project onto the linear combinations of invariant amplitudes, which reduce to the different gluon GPDs in the light cone limit and enable their separation in a lattice QCD calculation for spin-$0$ and spin-$\frac{1}{2}$ hadrons. Hence, this work lays the foundation for the numerical determination of the gluon GPDs from first-principle lattice QCD calculations, directly advancing our understanding of the mass and spin structures and mechanical properties of the nucleon, as well as the physics underlying deeply virtual Compton scattering and deeply virtual meson production in a range of experimental processes.  
\end{abstract}

\maketitle

%%%%%

\section{Introduction} \label{sec:intro1}

Generalized parton distributions (GPDs)~\cite{Muller:1994ses,Ji:1996ek,Radyushkin:1997ki} play a crucial role in providing a comprehensive description of the internal structure of hadrons and various emergent properties that arise due to the confining motion of quark and gluons inside the hadrons. For example, the moments of GPDs encode emergent properties such as the hadron's mass~\cite{Ji:1994av,Ji:1995sv} and spin~\cite{Ji:1996ek} structures.  GPDs also provide valuable insights into the orbital angular momentum of quarks and gluons, a crucial component in addressing the proton spin puzzle~\cite{EuropeanMuon:1987isl,EuropeanMuon:1989yki}.

Moreover, a deeper understanding of GPDs provides access to the pressure and shear forces that define the internal structure and dynamics of hadrons~\cite{Polyakov:2002wz,Polyakov:2002yz,Polyakov:2018zvc}. In particular, the second moment of GPDs is connected to the gravitational form factors (GFFs), which are the form factors of the quantum chromodynamics (QCD) energy-momentum tensor and contain important information about the mass, angular momentum, and mechanical properties of the nucleon~\cite{Polyakov:2018zvc,Lorce:2018egm}. (For the first experimental and lattice QCD determinations of mechanical properties of hadrons, see~\cite{Burkert:2018bqq,Shanahan:2018nnv}).

Unlike  parton distribution functions (PDFs), which give the probability of finding a parton (quark or gluon) carrying a specific fraction of the hadron's momentum, GPDs encode additional information about the spatial distributions and correlations of these partons. Thus, in contrast to the momentum fraction $x$ dependence of the PDFs, GPDs are a function of three variables: longitudinal momentum fraction $x$, fraction of longitudinal transferred momentum or skewness $\xi$, and the momentum transfer squared between the initial and final hadron that is proportional to the invariant momentum transfer $t \equiv \Dl^2$. In the zero skewness limit, Fourier transforming GPDs with respect to the transverse momentum transfer $\Dl_T$ yields the impact-parameter-dependent distributions $f(x,{\bf b}^2_T)$~\cite{Burkardt:2000za}. This makes GPDs essential for understanding the three-dimensional structure of hadrons, combining both longitudinal momentum and transverse spatial distributions~\cite{Ralston:2001xs, Diehl:2002he}.

GPDs are experimentally accessible through the deep virtual Compton scattering(DVCS)~\cite{Ji:1996nm,Radyushkin:1996ru,Radyushkin:1997ki,Collins:1996fb} and deeply virtual meson production~\cite{Goeke:2001tz,Diehl:2003ny,Vanderhaeghen:1999xj}, timelike Compton scattering~\cite{Berger:2001xd},  double DVCS~\cite{Guidal:2002kt,Belitsky:2002tf}, diphoton production~\cite{Grocholski:2022rqj}, photon meson pair production~\cite{Duplancic:2018bum}, diffractive leptoproduction of vector mesons~\cite{Brodsky:1994kf}, and single diffractive hard exclusive processes~\cite{Qiu:2022pla}. Near-threshold heavy quarkonium  $J/\psi$ production has been carried out by experiments at Jefferson Laboratory~\cite{GlueX:2019mkq,Duran:2022xag,GlueX:2023pev} to elucidate the role of GPDs and GFFs in the nucleon structure. Accessing GPDs is also a key focus for the upcoming Electron-Ion Collider (EIC)~\cite{Accardi:2012qut,AbdulKhalek:2021gbh} at Brookhaven National Laboratory and the Electron-Ion Collider in China~\cite{Anderle:2021wcy}.

In exclusive heavy meson production, the gluon GPDs are important~\cite{Goloskokov:2006hr,Goloskokov:2009ia,Goloskokov:2011rd} and it has been shown that heavy quarkonium production amplitude can be factorized in terms of gluon GPDs and  quarkonium distribution amplitude~\cite{Ivanov:2004vd,Guo:2021ibg}. Additionally, the single-spin asymmetry and double-spin asymmetry can be measured in heavy vector meson production ($ep \to epV$), both of which can be used to study the GPD $E_g$~\cite{Koempel:2011rc,Goloskokov:2024egn}. A recent publication~\cite{Guo:2023qgu} shows that the Compton-like amplitudes associated with $J/\psi$ production near the threshold are related to gluon GPDs at large $\xi$.

Regardless of the target hadron's spin, the experimentally accessible channels alone do not permit an unambiguous extraction of GPDs~\cite{Bertone:2021yyz} and extracting various GPDs can be challenging~\cite{Moffat:2023svr}. Additionally, the limited availability of experimental data and a small number of experimentally accessible channels sensitive to gluonic structures pose significant challenges to isolating the various gluon GPDs.  It is therefore critically important to determine the nonperturbative gluon GPDs from the first-principle lattice QCD calculations~\cite{Ji:2013dva} as highlighted in~\cite{Abir:2023fpo}.   We emphasize that a Lorentz-covariant parameterization of the matrix elements in terms of a linearly independent basis of tensor structures is crucial for projecting lattice QCD matrix elements onto light cone gluon GPDs. However, as we demonstrate below, the computation of off-forward gluonic matrix elements is significantly more complicated, and projections of lattice QCD matrix elements onto light cone gluon GPDs has not yet been explored in the literature to our knowledge.  This article aims to identify the off-forward gluonic matrix elements required to isolate and determine all the eight gluon GPDs ($H_g$, $E_g$, $\wt{H}_g$, $\wt{E}_g$, $H^T_g$, $E^T_g$, $\wt{H}^T_g$, $\wt{E}^T_g$ ) that can be used to perform numerical calculations using the quasi-PDFs and large momentum effective theory (LaMET)~\cite{Ji:2013dva, Ji:2014gla} and also the pseudo-PDF~\cite{Radyushkin:2017cyf} formalisms.

\section{Physics goals and essence of lattice QCD calculation of gluon GPDs} \label{sec:gluonGPDs}
Similar to the quark GPDs, there are eight leading-twist gluon GPDs for the nucleon. Among them are $H_g,~E_g,~\wt{H}_g,~\wt{E}_g$  and the gluon transversity GPDs  $H^T_g,~E^T_g,~\wt{H}^T_g ,~\wt{E}^T_g$~\cite{Hoodbhoy:1998vm,Diehl:2001pm}.  In ordinary PDFs, gluon helicity flip can only occur for
targets with spin-$1$ or higher~\cite{Jaffe:1989xy,Artru:1989zv}, since the change of helicity on the parton
side must be compensated by a corresponding change for the target in order
to ensure angular momentum conservation.  Therefore, there is no gluon transversity PDF for the nucleon. However, there is no such constraint
for GPDs, because they involve transverse
momentum transfer and thus  generalized gluon
helicity-flip distributions for a spin-$\frac{1}{2}$ target  involves orbital angular momentum transfer between partons.

Different GPDs provide essential insights into the structure of hadrons. For example, $E_g$ is related to the distortion of the unpolarized gluon distribution in a transversely polarized proton and can be related to the gluon Sivers function which is a T-odd transverse momentum-dependent PDF~\cite{Meissner:2007rx}. $J/\psi$ spin asymmetries are determined by $E_g$ and $\wt{H}_g$. However, the determination of the $E_g$ is quite challenging~\cite{Aschenauer:2013hhw} and the distribution of $E_g$  GPD is virtually unknown.

Calculating $H_g(x,\xi,t)$ and $E_g(x,\xi,t)$ gluon GPDs  give access to the total angular momentum contribution of the gluons to the proton spin $J_g$~\cite{Ji:1996ek}:
\bea
J_g = \frac{1}{2} \int dx~x~[H_q(x,0,0)+E_g (x,0,0)].
\eea
 $H_g$, $E_g$ and $\wt{H}_g$ GPDs can also be used to determine the gluon orbital angular momentum (OAM) in the nucleon~\cite{Ji:1996ek,Leader:2013jra}:
\bea
L^z_g = \frac{1}{2}\int dx~[x(H_g(x,0,0)+ E_g(x,0,0))-\wt{H}_g(x,0,0)].
\eea
This definition corresponds to the kinetic OAM of the gluon~\cite{Chen:2008ag,Wakamatsu:2010qj}. Additionally, in the
Jaffe-Manohar sum rule~\cite{Jaffe:1989jz}, the canonical gluon OAM distribution $\mathcal{L}_g(x)$ is related to the $E_g$ as~\cite{Hatta:2012cs}:
\bea
\mathcal{L}_g(x) &=& x \int_x^1 \frac{dx'}{x'} [H_g(x',0,0)+E_g(x',0,0)] - 2x\int_x^1 \frac{dx'}{{x'}^2}\Dl g(x') + \cdots ,
\eea
where $\Dl g(x)$ is the gluon helicity PDF and the genuine
twist-three terms are omitted. Understanding of  $E_g$ and its small-$x$ evolution is critically important for the contribution of $\mathcal{L}_g(x)$ in the nucleon spin budget~\cite{Hatta:2022bxn}. Gluon GPDs can shed light on the transverse spatial distributions of the gluons and their contribution to the total angular momentum and their spin-orbit correlations in the nucleon~\cite{Boer:2011fh,Accardi:2012qut,Achenbach:2023pba,Abir:2023fpo}.

Additionally, the nucleon being a composite system, the knowledge of the virtually unknown $E_g$ GPD is very important to understand the anomalous gravitomagnetic moment, which requires the gravitational form factor $B(Q^2)=0$ when summed over all quarks and gluons in the nucleon~\cite{Kobzarev:1962wt,Kobsarev:1970qm,Brodsky:2000ii}. This is equivalent to
\bea
\sum_{i=q,g}\int_{-1}^1 dx E^i(x,0,0)=0,
\eea
or the contributions from $B_{q,g}(0)$ to the total angular momentum $J_z$ cancel in the sum.

It has been proposed that the photoproduction of $J/\psi$ can be used to measure the gluon matrix element $\bra {P} F^2\ket{P}$ in the nucleon and provide information about the trace anomaly contribution to the nucleon mass~\cite{Kharzeev:1995ij,Kharzeev:1998bz}.  Knowledge of gluon GPDs can illuminate the physics of the threshold $J/\psi$ production and the form factors of the gluon energy-momentum tensor, providing important information on the mass structure of the nucleon. Understanding the nucleon mass is one of the major goals of the EIC physics program and lattice QCD calculations~\cite{Yang:2018nqn,He:2021bof,Wang:2024lrm}. Various  transversity GPDs provide information on the transverse polarization of gluons in transversely and longitudinally polarized nucleons. They also provide information on the 3D spin-orbit picture and can give insights into the connections between transverse polarization and orbital angular momentum~\cite{Diehl:2003ny}.

Additionally, understanding the gluonic structure of the pion is of particular theoretical interest, as the pion is the lightest QCD bound state and the Goldstone mode is associated
with dynamical chiral symmetry breaking. The gluon GPD of the pion remains essentially unknown from experiments, yet understanding it holds significant phenomenological implications for hadronic physics~\cite{Amrath:2008vx,Chavez:2021llq,deTeramond:2021lxc}. Because of the special
role of the pion in QCD, there have been sustained efforts  to explore its gluon distribution,
along with that of the nucleon, one of the main goals of
the upcoming EIC~\cite{Accardi:2012qut}.

Despite the above-mentioned fundamental importance of gluon GPDs, knowledge of various gluon GPDs is rather unknown~\cite{Diehl:2003ny,Polyakov:2002yz} from experiments. Since gluon GPDs are accessible in DVCS only at higher orders in $\alpha_s$ it is a difficult task to extract gluon GPDs from the data. Isolating different gluon GPDs from experimental data also poses a significant challenge~\cite{Kriesten:2021sqc}. Moreover,  transversity gluon GPDs are more difficult to probe experimentally than non-helicity-flip gluon GPDs (although there have been some suggestions, e.g. \cite{Pire:2017yge, Pire:2021dad}). Additionally, reconstructing the $x$ dependence of GPDs by measuring the $Q^2$ dependence of exclusive processes at fixed $\xi$ is challenging~\cite{Accardi:2012qut,Qiu:2023mrm} and presents a significant difficulty in extracting information from experimental data.

In contrast to recent progress in the lattice QCD (LQCD) calculations of various quark GPDs~\cite{Chen:2019lcm,Alexandrou:2019ali,Alexandrou:2020zbe,Lin:2020rxa,Lin:2021brq,Alexandrou:2021bbo,Bhattacharya:2022aob,Lin:2023gxz,Holligan:2023jqh,Bhattacharya:2023nmv,Bhattacharya:2024qpp,Hannaford-Gunn:2024aix,HadStruc:2024rix},  a few calculations of the unpolarized gluon PDFs in the nucleon, pion, and kaon~\cite{Fan:2020cpa,Fan:2021bcr,Fan:2022kcb,Salas-Chavira:2021wui,HadStruc:2021wmh,Delmar:2023agv,Good:2023ecp,Good:2024iur}, as well as matrix element calculations toward the gluon helicity Ioffe-time distribution~\cite{HadStruc:2022yaw}, and the first LQCD determination of the gluon helicity PDF~\cite{Khan:2022vot}, there has not been any LQCD calculations of the gluon GPDs.  In fact, a formalism based on the Lorentz-covariant parameterization of the Euclidean off-forward gluon matrix elements in terms of Lorentz-invariant amplitudes and their appropriate projection onto various light-cone gluon GPDs have not been performed in the literature. As the number of independent observables in experiments is typically insufficient to fully isolate and constrain all eight gluon GPDs,  LQCD calculations of all these gluon GPDs are of tremendous importance. The goal of this paper is to identify a set of projection matrices, to be applied to the spacelike separated bilocal gluonic operator, that correspond to each one of the eight leading-twist gluon GPDs respectively. Moreover, the projections shall be ``exact'' in the sense that there are no so-called ``contaminations". This concept will be introduced more clearly in Sec.~\ref{sec: spin0}. While this is not strictly necessary, it is useful, because the $z^2$ dependence of the Lorentz-invariant amplitudes are then ``disentangled'' from contaminations and may be studied explicitly on the lattice. 
This strategy was first introduced in \cite{Balitsky:2019krf} for the forward unpolarized gluon PDF and is here extended to the off-forward case for generic polarizations. As a side remark, we will also demonstrate that such projections are not unique in the sense that they depend on the basis chosen for the Lorentz decomposition of the matrix element. This is of particular relevance in the off-forward case, where the set of all possible Lorentz structures is generically overcomplete, which leads to an ambiguity in the basis choice.

%%%%%%%%%%%%%%%%%%%%
%%%%%%%%%%%%%%%%%%%%
\section{Definitions and kinematics} \label{sec:prelim}

In principle, all the information about gluon GPDs is contained within the following off-forward quasi-GPD hadronic matrix element~\cite{Ji:2013dva}:
\bea
    M_{s's}^{\mu \nu ; \alpha \beta}(z) = \bra{p',s'}  G^{\mu \nu}(-z/2) W(-z/2,z/2) G^{\alpha \beta}(z/2) \ket {p,s}.
    \label{eq: ME main}
\eea
The momenta $p,p'$, and spin projections $s,s'$ of the incoming and outgoing hadrons respectively are different in general. In the rest of the paper, we use the Minkowski spacetime notation. We use the following definitions:
\begin{align}
P = \frac{p+p'}{2}, \qquad \Delta = p' - p, \qquad m^2 = p^2 = {p'}^2, \qquad t = \Delta^2.
\end{align}
Formally, we always assume that we are in a frame where the hadron is highly boosted in the $3$ direction. Denoting by $\hat z^{\mu} = (0,0,0,1)^{\mu}$, the unit vector in the $3$ direction, this means that
\begin{align}
P_3 = \hat z \cdot P \gg m^2, -t.
\end{align}
Note that $t < 0$ kinematically.
Furthermore, we introduce the light cone vectors
\begin{align}
n^{\mu} = \frac{1}{2} ( \hat t^{\mu} + \hat z^{\mu} ), \qquad \bar n^{\mu} = \hat t^{\mu} - \hat z^{\mu},
\end{align}
where $\hat t^{\mu} = (1,0,0,0)^{\mu}$ is the unit vector in the time direction. 

A four-vector $v^{\mu}$ can be decomposed in terms of the light cone basis with respect to $n$ and $\bar n$ by
\begin{align}
v^{\mu} = v^+ \bar n^{\mu} + v^- n^{\mu} + v_{\perp}^{\mu},
\end{align}
where $v^+ = n \cdot v$, $v^- = \bar n \cdot v$. The choice of the factor $\frac{1}{2}$ in the definition of $n$ is convenient because it ensures that $v^+ = v_0 = v_3$ whenever $v^- = 0$.

We can further specify the frame by taking $P_{\perp}^{\mu} = (0, - P_1 , - P_2 ,0)^{\mu} = 0$. The only transverse vector (given that $z_{\perp}$ will  always be set to zero) is then $\Delta_{\perp}$, whose length is fixed by
\begin{align} \label{eq:delperp}
\sqrt{ - \Delta_{\perp}^2 }= \sqrt{ - 4m^2 \xi^2 - t (1-\xi^2) },
\end{align}
where we have defined the skewness $\xi$ by
\begin{align}
\xi = - \frac{\Delta^+}{2P^+}.
\label{eq: xi def}
\end{align}
The frame is then entirely specified, up to boosts in the $3$ direction, by taking $\Delta_{\perp}^{\mu}$ in the $1$ direction, i.e. $\Delta_1 = \sqrt{ - \Delta_{\perp}^2 }$. The off-forward generalization of the Ioffe time~\cite{Gribov:1965hf,Ioffe:1969kf,Braun:1994jq} involves the average of the nucleon momenta
\begin{align}
\omega = - z \cdot P.
\end{align}
It will also be convenient to define the so-called quasi-skewness \cite{Braun:2024snf}
\begin{align}
\eta = - \frac{z \cdot \Delta}{2z \cdot P} = \frac{z \cdot \Delta}{2\omega}.
\end{align}
We will consider two cases with respect to the properties of the four-vector $z$:
\begin{itemize}

    \item Case 1: $z \propto n$. Then we let
    \begin{align}
    z^{\mu} = z^- n^{\mu}, \qquad \omega = - z^- P^+.
    \end{align}
    $M(z)$ is the light cone distribution that is related by Fourier transformation to the GPDs (after performing a nonmultiplicative UV renormalization). $M$ can then depend on four Lorentz invariants: $\omega$, $\eta$, $t$, and $m^2$, in addition to nonperturbatively generated scales such as $\Lambda_{\rm QCD}$. Note that for the light cone kinematics, i.e. $z \propto n$, $\eta = \xi$. But $\xi$, as defined in Eq.\, \eqref{eq: xi def}, is strictly speaking not  Lorentz invariant (it is only invariant under boosts in the $3$ direction). 
\item Case 2: $z \propto \hat z$. Then we let
\begin{align}
z^{\mu} = - \sqrt{-z^2} \hat z^{\mu} = - z_3 \hat z^{\mu}, \qquad \omega = z_3 P_3,
\end{align}
where $z_3 = \sqrt{-z^2}$.
It is important to note that the $P_3 \rightarrow{\infty}$ limit at fixed $\omega \sim 1$ corresponds to the limit $z_3 \rightarrow 0$.
For this spatial vector $z$, $M(z)$ can be calculated on the lattice and matched perturbatively to the light cone distribution. $M$ depends on five Lorentz invariants which we choose as $\omega, \eta, t , m^2, z^2$. Again, $\xi$ can be written in terms of Lorentz invariants under the constraint that $z \propto \hat z$ as
\begin{align}
\xi = \frac{\eta}{\sqrt{1 -  \frac{m^2 z^2 }{\omega^2} + \frac{ t z^2 }{4\omega^2}}} = 
\eta + \mathcal O(P_3^{-2}).
\end{align}
We will generally assume  the case $z \propto \hat z$ in the subsequent calculation unless otherwise indicated. 
\end{itemize}

In the following, we begin by outlining the theoretical technique for determining the LQCD matrix elements that can be projected onto the light cone gluon GPD for spin-$0$ hadrons. We then extend this strategy to address the more complicated case of spin-$\frac{1}{2}$ hadrons.  To facilitate some of the results derived in this work,  we use {\it Mathematica} and FORM.

\section{Projection for the spin-$0$ hadron}
\label{sec: spin0}

For simplicity, we begin by focusing on the case of  a scalar (spin-$0$) hadron. 
We can then drop the spin projection indices in Eq.\,\eqref{eq: ME main}. By Lorentz covariance, $M^{\mu \nu; \alpha \beta}$ must be generated by tensors that can be built out of $g^{\mu \nu}, \varepsilon^{\mu \nu \rho \sigma}$ and the three available Lorentz vectors $P^{\mu}, \Delta^{\mu}, z^{\mu}$. Parity symmetry further implies that the Levi-Civita tensor $\varepsilon^{\mu \nu \rho \sigma}$ cannot appear. This applies regardless of whether the hadron is a scalar or pseudoscalar, as the in- and outgoing hadronic states in the matrix element have the same intrinsic parity, causing any possible phase to cancel. Thus we have to include in a decomposition 
\begin{align}
M^{\mu \nu; \alpha \beta}(\omega, \eta, t, z^2) = \sum_{\ell} \mathcal M_{\ell}(\omega, \eta, t, z^2) \mathcal T_{\ell}^{\mu \nu; \alpha \beta}(\omega, \eta, t, z^2)
\label{eq: Ansatz scalar}
\end{align}
all the possible tensors that can be built out of the metric $g^{\mu \nu}$ and $P^{\mu}, \Delta^{\mu}, z^{\mu}$. 

In Eq.~\eqref{eq: Ansatz scalar} the $\mathcal M_{\ell}$ are functions of the Lorentz invariants $\omega, \eta, t, m^2, z^2$.  Note that the $\mathcal M_{\ell}$'s suffer from UV divergences and they must be renormalized, resulting in a dependence on a renormalization scale $\mu$. The nature of these UV divergences changes depending on $z^2$. Generally one can expect multiplicative renormalization for $z^{\mu} \propto \hat z^{\mu}$ and nonlocal (convolutive) renormalization for $z^{\mu} \propto n^{\mu}$. We do not discuss the issue of UV divergences in this work and we always tacitly assume that $\mathcal M_{\ell}$ are appropriately renormalized in some way. For definiteness, we take that $\mathcal M_{\ell}$ at $z^2 = 0$ to be renormalized in the $\overline{\text{MS}}$ scheme with the usual nonlocal GPD kernels.
The renormalization procedure for $z^2 < 0$ is standard~\cite{Yao:2022vtp} using the ratio~\cite{Radyushkin:2017cyf} and hybrid renormalization~\cite{Ji:2020brr} schemes and will be a subject for future study with the determination of appropriate matching kernels within the LaMET and short-distance factorization based approaches.

Note that further symmetries of the matrix element in Eq.\, \eqref{eq: ME main}, e.g. permuting the gluon operators $(\mu, \nu, z) \leftrightarrow (\alpha, \beta, -z) $ and time reversal, imply relations among the $\mathcal M_{\ell}$. Since this does not influence the linear (in)dependence of the tensors, we will not comment on such symmetries further.

The central assumption is that each $\mathcal M_{\ell}$ admits a short-distance factorization of the following schematic form
\begin{align}
\mathcal M_{\ell}(\omega, \eta, t, z^2) = \sum_{f} C_{\ell f}(\log (-\mu_F^2z^2), \omega, \eta) \otimes \mathcal M_{f}(\omega, \eta, t, z^2 = 0, \mu_F) + \text{power corrections},
\label{eq: matching general}
\end{align}
where $\otimes$ denotes some convolution product and $C_{\ell f} = \delta_{\ell f} \mathbb{I} + O(\alpha_s)$ is a perturbative matching kernel and we have displayed the dependence on the factorization scale $\mu_F$ for this equation only. Note that the matching kernel mixes not only invariant amplitudes of $M$ but it also induces mixing with matrix elements of other operators, containing quarks, say. Equation\,\eqref{eq: matching general} is subject to power corrections that vanish in the limit $z_3 \rightarrow 0$, an important subject for ongoing and future investigations. In the context of GPDs there is some knowledge of the functional form from the study of renormalons \cite{Braun:2024snf}. We remark that, in the GPD case, for generic components of $M$, there are also twist-three $\sim 1/P_3$ corrections. For instance, there are so-called ``kinematic" twist-three corrections that are required to restore translational invariance, which have so far been investigated for the quark case \cite{Braun:2023alc}. We do not discuss power corrections further in this work.

The list of possible tensors that may occur in the decomposition Eq.\,\eqref{eq: Ansatz scalar} reads
\begin{align} \notag
\mathcal{T}_{gg}&=g^{\alpha  \nu } g^{\beta  \mu }-g^{\alpha  \mu } g^{\beta  \nu },
\\ \notag
\mathcal{T}_{PP}&=P^{\alpha } P^{\mu } g^{\beta  \nu }-P^{\beta } P^{\mu } g^{\alpha  \nu }-P^{\alpha } P^{\nu } g^{\beta  \mu }+P^{\beta } P^{\nu } g^{\alpha  \mu },
\\ \notag
\mathcal{T}_{P\Delta}&=P^{\alpha } \Delta ^{\mu } g^{\beta  \nu }-P^{\beta } \Delta ^{\mu } g^{\alpha  \nu }-P^{\alpha } \Delta ^{\nu } g^{\beta  \mu }+P^{\beta } \Delta ^{\nu } g^{\alpha  \mu },
\\ \notag
\mathcal{T}_{\Delta P}&=P^{\mu } \Delta ^{\alpha } g^{\beta  \nu }-P^{\nu } \Delta ^{\alpha } g^{\beta  \mu }-P^{\mu } \Delta ^{\beta } g^{\alpha  \nu }+P^{\nu } \Delta ^{\beta } g^{\alpha  \mu },
\\ \notag
\mathcal{T}_{\Delta \Delta}&=\Delta ^{\alpha } \Delta ^{\mu } g^{\beta  \nu }-\Delta ^{\beta } \Delta ^{\mu } g^{\alpha  \nu }-\Delta ^{\alpha } \Delta ^{\nu } g^{\beta  \mu }+\Delta ^{\beta } \Delta ^{\nu } g^{\alpha  \mu },
\\ \notag
\mathcal{T}_{PP\Delta \Delta} &=P^{\beta } P^{\nu } \Delta ^{\alpha } \Delta ^{\mu }-P^{\alpha } P^{\nu } \Delta ^{\beta } \Delta ^{\mu }-P^{\beta } P^{\mu } \Delta ^{\alpha } \Delta ^{\nu }+P^{\alpha } P^{\mu } \Delta ^{\beta } \Delta ^{\nu },
\\ \notag
\mathcal{T}_{Pz}&=P^{\alpha } z^{\mu } g^{\beta  \nu }-P^{\beta } z^{\mu } g^{\alpha  \nu }-P^{\alpha } z^{\nu } g^{\beta  \mu }+P^{\beta } z^{\nu } g^{\alpha  \mu },
\\ \notag
\mathcal{T}_{\Delta z}&=\Delta ^{\alpha } z^{\mu } g^{\beta  \nu }-\Delta ^{\alpha } z^{\nu } g^{\beta  \mu }-\Delta ^{\beta } z^{\mu } g^{\alpha  \nu }+\Delta ^{\beta } z^{\nu } g^{\alpha  \mu },
\\ \notag
\mathcal{T}_{PP\Delta z}&=P^{\beta } P^{\nu } \Delta ^{\alpha } z^{\mu }-P^{\beta } P^{\mu } \Delta ^{\alpha } z^{\nu }-P^{\alpha } P^{\nu } \Delta ^{\beta } z^{\mu }+P^{\alpha } P^{\mu } \Delta ^{\beta } z^{\nu },
\\ 
\mathcal{T}_{P\Delta \Delta z}&=-P^{\beta } \Delta ^{\alpha } \Delta ^{\mu } z^{\nu }+P^{\alpha } \Delta ^{\beta } \Delta ^{\mu } z^{\nu }+P^{\beta } \Delta ^{\alpha } \Delta ^{\nu } z^{\mu }-P^{\alpha } \Delta ^{\beta } \Delta ^{\nu } z^{\mu },
\label{eq: tensor list}
\\ \notag
\mathcal{T}_{zP}&=P^{\mu } z^{\alpha } g^{\beta  \nu }-P^{\nu } z^{\alpha } g^{\beta  \mu }-P^{\mu } z^{\beta } g^{\alpha  \nu }+P^{\nu } z^{\beta } g^{\alpha  \mu },
\\ \notag
\mathcal{T}_{z\Delta}&=\Delta ^{\mu } z^{\alpha } g^{\beta  \nu }-\Delta ^{\mu } z^{\beta } g^{\alpha  \nu }-\Delta ^{\nu } z^{\alpha } g^{\beta  \mu }+\Delta ^{\nu } z^{\beta } g^{\alpha  \mu },
\\ \notag
\mathcal{T}_{PPz \Delta}&=P^{\beta } P^{\nu } \Delta ^{\mu } z^{\alpha }-P^{\alpha } P^{\nu } \Delta ^{\mu } z^{\beta }-P^{\beta } P^{\mu } \Delta ^{\nu } z^{\alpha }+P^{\alpha } P^{\mu } \Delta ^{\nu } z^{\beta },
\\ \notag
\mathcal{T}_{P\Delta z \Delta}&=-P^{\nu } \Delta ^{\alpha } \Delta ^{\mu } z^{\beta }+P^{\nu } \Delta ^{\beta } \Delta ^{\mu } z^{\alpha }+P^{\mu } \Delta ^{\alpha } \Delta ^{\nu } z^{\beta }-P^{\mu } \Delta ^{\beta } \Delta ^{\nu } z^{\alpha },
\\ \notag
\mathcal{T}_{zz}&=z^{\alpha } z^{\mu } g^{\beta  \nu }-z^{\beta } z^{\mu } g^{\alpha  \nu }-z^{\alpha } z^{\nu } g^{\beta  \mu }+z^{\beta } z^{\nu } g^{\alpha  \mu },
\\ \notag
\mathcal{T}_{PPzz}&=P^{\beta } P^{\nu } z^{\alpha } z^{\mu }-P^{\alpha } P^{\nu } z^{\beta } z^{\mu }-P^{\beta } P^{\mu } z^{\alpha } z^{\nu }+P^{\alpha } P^{\mu } z^{\beta } z^{\nu },
\\ \notag
\mathcal{T}_{P\Delta zz}&=-P^{\beta } \Delta ^{\mu } z^{\alpha } z^{\nu }+P^{\alpha } \Delta ^{\mu } z^{\beta } z^{\nu }+P^{\beta } \Delta ^{\nu } z^{\alpha } z^{\mu }-P^{\alpha } \Delta ^{\nu } z^{\beta } z^{\mu },
\\ \notag
\mathcal{T}_{\Delta P zz}&=-P^{\nu } \Delta ^{\alpha } z^{\beta } z^{\mu }+P^{\mu } \Delta ^{\alpha } z^{\beta } z^{\nu }+P^{\nu } \Delta ^{\beta } z^{\alpha } z^{\mu }-P^{\mu } \Delta ^{\beta } z^{\alpha } z^{\nu },
\\ \notag
\mathcal{T}_{\Delta \Delta zz}&=\Delta ^{\alpha } \Delta ^{\mu } z^{\beta } z^{\nu }-\Delta ^{\beta } \Delta ^{\mu } z^{\alpha } z^{\nu }-\Delta ^{\alpha } \Delta ^{\nu } z^{\beta } z^{\mu }+\Delta ^{\beta } \Delta ^{\nu } z^{\alpha } z^{\mu }.
\end{align}
We emphasize that contrary to the forward case \cite{Balitsky:2019krf}, where the tensors involving $\Delta$ are absent, the set of these tensors is not linearly independent. Given the absence of additional insights,
potential linear relations between the tensors can be derived systematically by using the Gram-Schmidt procedure with respect to the inner product
\begin{align}
(u,v) = u_{\mu \nu; \alpha \beta} v^{\mu \nu; \alpha \beta},
\end{align}
where $u,v$ are linear combinations of the $\mathcal T_{\ell}$ in Eq.\,\eqref{eq: tensor list}. 
Note that, while this inner product is linear and symmetric, it is indefinite. In particular, $(u,u) = 0$ does not imply $u = 0$. Thus, if we encounter a $u$ that has $(u,u) = 0$ in the Gram-Schmidt algorithm, we need to check whether $u$ is actually zero. In this way, one can find a single linear relation
\begin{align} \notag
0 &= -\Delta _1^2 \eta ^2 \omega ^2 \mathcal{T}_{gg} +\left(4 \xi ^2 \omega ^2-\Delta _1^2
   z_3^2\right) \xi ^2 \mathcal{T}_{PP} +2 \eta  \xi ^2 \omega ^2 ( \mathcal{T}_{P\Delta} + \mathcal{T}_{\Delta P}) +\eta ^2 \omega ^2 
   \mathcal{T}_{\Delta \Delta}  -\xi ^2 z_3^2 \mathcal{T}_{PP\Delta \Delta}
\\ \label{eq: linrel}
&\quad -\xi ^2 t   \omega\mathcal{T}_{Pz}+2 \eta  \xi ^2 P^2 \omega \mathcal{T}_{\Delta z}  -2 \eta  \xi ^2\omega  \mathcal{T}_{PP\Delta z} -\xi ^2 \omega
   \mathcal{T}_{P\Delta \Delta z}  -\xi ^2 t \omega \mathcal{T}_{zP}  +2 \eta  \xi ^2 P^2 \omega \mathcal{T}_{z\Delta} 
\\ \notag
&\quad -2 \eta  \xi ^2  \omega\mathcal{T}_{PPz \Delta}  -\xi ^2 \omega\mathcal{T}_{P\Delta z \Delta} 
   -\xi ^2 P^2 t \mathcal{T}_{zz} +\xi ^2 t
   \mathcal{T}_{PPzz}
    +\xi ^2 P^2
   \mathcal{T}_{\Delta \Delta zz},
\end{align}
which holds for $z^2 < 0$ as well as $z^2 = 0$. Written in this way, the relation is nontrivial only if $\xi, \Delta_1, \omega \neq 0$. After dividing by $\xi^2$, we can take the limit $\xi \rightarrow 0$, so we obtain a nontrivial relation also at $\xi = 0$. Note that the situation is quite different if $\Delta_1 = 0$, in which case $\Delta$ is linearly dependent on $P$ and $z$, so the tensors involving $\Delta$ in Eq.~\eqref{eq: tensor list} can be dropped so that the remaining tensors form a linearly independent set. Then we have the same basis as in the forward case \cite{Balitsky:2019krf}.

We focus for now on the case where $\Delta_1 \neq 0$. We can then eliminate by the virtue of Eq.\,\eqref{eq: linrel} one tensor from Eq.\,\eqref{eq: tensor list} to obtain a basis of 18 elements. The generalized Ioffe-time distribution (GITD) $\frak F(\omega, \xi, t)$ is then determined in terms of the corresponding invariant amplitudes $\mathcal M_{\ell}|_{z^2 = 0}$ and coefficients $f_{\ell} = f_{\ell}(\omega, \eta, t)$ through
\begin{align}
\frak F = z_{\mu} z_{\alpha} g_{\nu \beta} M^{\mu \nu ; \alpha \beta}  |_{z^2 = 0} = \sum_{\ell} f_{\ell}  \, \mathcal M_{\ell} |_{z^2 = 0}.
\end{align}
Note that the Fourier transform of $\frak F$ with respect to $\omega$ gives a GPD of a spin-$0$ particle
\begin{align}
H_{g/h}(x,\xi,t) = \int_{-\infty}^{\infty} d\omega \, e^{i\omega x} \frak F(\omega, \xi, t),
\end{align}

For instance, if we eliminate the $\mathcal T_{\Delta \Delta}$ structure, we obtain
\begin{align}
\frak F = -2\omega^2 \mathcal M_{PP} |_{z^2 = 0} + 4\xi \omega^2 \left (\mathcal M_{P\Delta}|_{z^2 = 0} + \mathcal M_{\Delta P}|_{z^2 = 0} \right ) + \omega^2 \Delta_1^2 \mathcal M_{PP \Delta \Delta}|_{z^2 = 0}, 
\end{align}
whereas if we eliminate the $\mathcal T_{gg}$ structure we obtain
\begin{align}
\frak F = -2\omega^2 \mathcal N_{PP}|_{z^2 = 0} + 4\xi \omega^2 \left (\mathcal N_{P\Delta}|_{z^2 = 0} + \mathcal N_{\Delta P}|_{z^2 = 0} \right ) + \omega^2 \Delta_1^2 \mathcal N_{PP \Delta \Delta}|_{z^2 = 0} - 8 \xi^2 \omega^2 \mathcal N_{\Delta \Delta}|_{z^2 = 0}.
\end{align}
We have used different symbols $\mathcal N$ instead of $\mathcal M$ in order to emphasize that the Lorentz-invariant amplitudes, which are defined to be the coefficients of the basis tensors transform accordingly when changing the basis (so that $M$ is invariant, see below), i.e. $\mathcal N_{\ell} \neq \mathcal M_{\ell}$ in general.

The ``projection problem" is now to find a projection $\mathcal P_{\mu \nu; \alpha \beta}$, so that
\begin{align}
(\mathcal P, M) = \mathcal P_{\mu \nu; \alpha \beta} M^{\mu \nu; \alpha \beta} = \sum_{\ell} f_{\ell} \mathcal M_{\ell}. 
\label{eq: proj def}
\end{align}
Note that $\mathcal P$ has to be a tensor, since $f_{\ell}$ and $\mathcal M_{\ell}$ are Lorentz invariant by construction. 
Furthermore, since we have removed the linearly dependent structure in Eq.\,\eqref{eq: tensor list}, the existence of $\mathcal P$ is guaranteed, since by construction $\det (\mathcal T_{\ell}, \mathcal T_{\ell'})_{\ell \ell'} \neq 0$. Clearly, the projection problem corresponds to solving a system of linear equations, whose solution can be obtained, for example, by inverting the matrix $(\mathcal T_{\ell}, \mathcal T_{\ell'})_{\ell \ell'}$.  If $\mathcal P$ does not solve the projection problem, the remainders $0 \neq (\mathcal P, M) -  \sum_{\ell} f_{\ell} \mathcal M_{\ell} = \mathcal O(z^2)$ are commonly referred to as contaminations.

However, since we only demand that the projected matrix element $\sum_{\ell} f_{\ell} \mathcal M_{\ell}$ agrees with the GITD at $z^2 = 0$, $\mathcal P$ is not unique in the sense that it depends on the basis. More precisely, two projections $\mathcal P, \, \mathcal P'$ that satisfy Eq.\,\eqref{eq: proj def} may differ by terms of $\mathcal O(z^2)$, that is
\begin{align}
(\mathcal P - \mathcal P', M ) = \mathcal O(z^2).
\end{align}
Such differences correspond to the change of basis. Consider a Lorentz-invariant ``rotation'' $R$ defined by
\begin{align}
\mathcal S_r = \sum_{\ell} R_{r\ell} \mathcal T_{\ell} \equiv (R \, \mathcal T)_r,
\end{align}
which must be invertible and be decomposed as 
\begin{align}
R = \bar R + \widehat R,
\end{align}
where $\bar R = \bar R(\omega, \eta, t , m^2)$ does not depend on $z^2$ and $\widehat R = \widehat R(\omega, \eta, t, m^2 , z^2) = \mathcal O(z^2)$. For example, choosing a different tensor to be eliminated by the virtue of Eq.\,\eqref{eq: linrel} corresponds to such a rotation. We have to demand that $M$ is invariant under such a rotation
\begin{align}
M &= \sum_{\ell} \mathcal M_{\ell} \mathcal T_{\ell} \quad ( \equiv \mathcal M^T \mathcal T ) = \sum_{r} \mathcal N_r \mathcal S_r \quad ( \equiv  \mathcal N^T \mathcal S ), 
\end{align}
so that $\mathcal N^T = \mathcal M^T R^{-1}$. Also $\mathfrak F$ must be invariant, which implies
\begin{align}
\mathfrak F &= \sum_{\ell} \mathcal M_{\ell} |_{z^2 = 0}  f_{\ell}  \quad ( \equiv \mathcal M^T |_{z^2 = 0} f ) = \sum_{r} \mathcal N_r |_{z^2 = 0} g_r \quad ( \equiv  \mathcal N^T|_{z^2 = 0} g ), 
\end{align}
so that $g = \bar R f$. Now, suppose $\mathcal P[\mathfrak F]$ and $\mathcal P'[\mathfrak F]$ solve the projection problem in the bases given by $\mathcal T_{\ell}$ and $\mathcal S_r$ respectively, that is
\begin{align}
(\mathcal P[\mathfrak F], M) = \mathcal M^T f , \qquad (\mathcal P'[\mathfrak F], M) = \mathcal N^T g. 
\label{eq: P P' def}
\end{align}
Then, we must have
\begin{align}
(\mathcal P[\mathfrak F] - \mathcal P'[\mathfrak F], M) = \mathcal M^T f - \mathcal M^T R^{-1} \bar R f = \mathcal M^T R^{-1} \widehat R f = \mathcal O(z^2).
\end{align}
Moreover, anytime we have a projection $\mathcal P$ that solves the projection problem up to $\delta \mathcal M \equiv z^2 \mathcal M^T h$, where $h = \mathcal O((z^2)^0)$, in some basis, i.e. $(\mathcal P, M) = \mathcal M^T f + \delta \mathcal M$, there exists a transformation of basis $R = \mathbb I + \mathcal O(z^2)$ such that $\mathcal P$ solves the projection problem in the new basis, i.e. $(\mathcal P, M) = \mathcal N^T f$. Indeed, it is easy to see that $R$ is given such that $(R - \mathbb I) (f+ z^2 h) = -z^2 h$.

We now discuss an explicit example. Let $\mathcal T_{\ell}$ denote the basis obtained by removing $\mathcal T_{\Delta \Delta}$ from Eq.\,\eqref{eq: tensor list} and $\mathcal S_r$ be the basis obtained by removing $\mathcal T_{gg}$. Further, let $\mathcal P[\frak F]$ and $\mathcal P'[\frak F]$ be the projections that solve the projection problem\,\eqref{eq: proj def} in the bases $\mathcal T_{\ell}$ and $\mathcal S_r$ respectively. The system of equations can be solved analytically, most conveniently by using a symbolical computer algebra program.
One obtains
\begin{align}
(\mathcal P[\frak F], M) &= \frac{z_3^2 \xi^2}{\eta^2} (M_{0i;0i} + M_{ij;ij}) + \frac{4\omega \xi^2 (\xi - \eta)  z_3 }{\eta^2 \Delta_1} (M_{02;12} + M_{12;02})
\label{eq: spin0 proj1}
\end{align}
and
\begin{align}
(\mathcal P'[\frak F], M) &=  \frac{z_3^2 \xi^2}{\eta^2} M_{0i;0i} + \frac{\xi^2 [16 \omega^2 \xi (\xi - \eta) + z_3^2(4\xi^2 P^2-t)]}{2\eta^2 \Delta_1^2} M_{ij;ij} + \frac{4 \omega \xi^2 (\xi - \eta) z_3 }{\eta^2 \Delta_1} (M_{02;12} + M_{12;02}).
\label{eq: spin0 proj2}
\end{align}
The difference between the projections is 
\begin{align}
(\mathcal P[\frak F] - \mathcal P'[\frak F], M) &= \frac{\xi^2}{\eta^3 \Delta_1^2} \left [16 \omega^2 \xi^2 (\xi - \eta) - z_3^2 \left (t \eta + 4P^2 \xi^2 (3\eta - 4\xi) \right ) \right ]\mathcal M_{gg} \quad \stackrel{\Delta \rightarrow 0}{\longrightarrow} \quad \frac{\omega^2}{p_0^2} \mathcal M_{gg}.
\end{align}
Note that while $\mathcal P[\frak F]$ reproduces the projection from \cite{Balitsky:2019krf} in the forward limit
\begin{align}
(\mathcal P[\frak F], M) \stackrel{\Delta \rightarrow 0}{\longrightarrow} \frac{\omega^2}{p_0^2} (M_{0i;0i} + M_{ij;ij}) = - 2 \omega^2 \mathcal M_{PP},
\end{align}
$\mathcal P'[\frak F]$ becomes
\begin{align}
(\mathcal P'[\frak F], M) \stackrel{\Delta \rightarrow 0}{\longrightarrow} \frac{\omega^2}{p_0^2} \left  ( M_{0i;0i} + \frac{1}{2} M_{ij;ij} \right ) = -2 \omega^2 \mathcal N_{PP} = -2 \omega^2 \left ( \mathcal M_{PP} - \frac{1}{p_0^2} \mathcal M_{gg} \right ).
\end{align}
We conclude that the change of basis given by eliminating $\mathcal T_{gg}$ rather than $\mathcal T_{\Delta \Delta}$ corresponds to absorbing the contaminating amplitude $\mathcal M_{gg}$ into $\mathcal M_{PP}$
\begin{align}
\mathcal M_{PP} \rightarrow \mathcal  N_{PP} = \mathcal M_{PP} - \frac{1}{p_0^2} \mathcal M_{gg} \qquad (\Delta = 0)
\end{align}
in the forward limit. 

The conclusion is that the projection problem does not provide a justification that certain projections ``reduce'' $\mathcal O(z^2)$ power corrections. Nevertheless, it is useful to identify, given a choice of basis, the linear combinations of invariant amplitudes that gives the light cone distribution at $z^2 = 0$,  so that their $z^2$ dependence may be studied explicitly on the lattice. For this we propose the basis obtained by eliminating $\mathcal T_{gg}$, leading to the projection $\mathcal P$ in Eq.\,\eqref{eq: spin0 proj1}, since on the one hand, the result is simpler than $\mathcal P'$ in Eq.\,\eqref{eq: spin0 proj2}, and on the other hand $\mathcal P$ reproduces the projection from \cite{Balitsky:2019krf} in the forward limit, obtained by taking the most simple ``canonical'' basis.

\section{Projection for the spin-$\frac{1}{2}$ hadron}

The case of the spin-$\frac{1}{2}$ hadron is much more complicated than that of the spin-$0$ hadron since one has an order of magnitude more Lorentz structures.  
To work out a Lorentz-covariant decomposition of the form
\begin{align}
M_{s's}^{\mu \nu; \alpha \beta} = \frac{1}{2m}\sum_{\ell} \mathcal M_{\ell} \mathcal T_{\ell \, s's}^{\mu \nu; \alpha \beta},
\label{eq: Ansatz spinhalf}
\end{align}
we can use various symmetries and Gordon identities, i.e. the equations of motion $(\slashed p - m) u_s(p) = \bar u_{s'}(p') (\slashed p' - m) = 0$. The factor $\frac{1}{2m}$ is conventional. As for the spinor structures $\bar u_{s'}(p') \Gamma u_s(p)$ (which must appear in each term exactly once), we can use the basis $\Gamma \in \{ 1, \gamma_5, \gamma^{\mu} , \gamma^{\mu} \gamma_5, \sigma^{\mu \nu} \}$.
A great simplification is due to the parity symmetry. Since $M$ is parity-even, the Levi-Civita tensor can only appear with $\bar u_{s'}(p') \{ \gamma_5, \gamma^{\mu} \gamma_5 \} u_s(p)$ and vice versa. However, implementing these symmetries in the decomposition is not strictly necessary, since we will eventually determine all linear relations between the $\mathcal T_{\ell}$'s systematically by the Gram-Schmidt procedure. 

Using only the Gordon identities, which are discussed in Appendix \ref{app: gordon}, together with the parity symmetry one can readily see that we can decompose $M_{s's}$ as
\begin{align}
M_{s's} = f_S \bar u_{s'}(p') u_s(p) + f_T \bar u_{s'}(p') i\sigma^{\overbrace{\ast \ast}^{(\text{no }P)}} u_s(p),
\label{eq: basis1}
\end{align}
where the $f$'s are tensors that can be built out of the available vectors $P^{\mu}, \Delta^{\mu}, z^{\mu}$ and the metric tensor $g^{\mu \nu}$. In particular, the Levi-Civita tensor can be entirely eliminated.
All indices can be arbitrarily contracted. This is, of course, under the condition that the four open indices of $M$  are open on the right-hand side of Eq.~\eqref{eq: basis1}. We have indicated that  contractions of the vector $P$ with $\sigma^{\ast \ast}$ can be excluded by virtue of the Gordon identity\,\eqref{eq: G2}.

Using furthermore only the antisymmetry of $M^{\mu \nu; \alpha \beta}$ under $\mu \leftrightarrow \nu$ and $\alpha \leftrightarrow \beta$ we find that there is a total of 102 different structures $\mathcal T_{\ell}$ that correspond to the form in Eq.\,\eqref{eq: basis1}. As in the spin-$0$ we must eliminate all remaining linear dependencies in order to make the invariant amplitudes $\mathcal M_{\ell}$ well defined.
For this we can proceed analogously, by defining an inner product
\begin{align}
(u,v) &= (u_{s's;\mu \nu; \alpha \beta})^* v_{s's}^{\mu \nu; \alpha \beta},
\end{align}
with summation of $s,s'$ implied and ${}^*$ denotes complex conjugation. Note that typically one obtains in a lattice calculation a matrix $\textbf B$ in the Dirac space defined by
\begin{align}
M_{s's}^{\mu \nu; \alpha \beta} = \bar u_{s'}(p') \textbf B^{\mu \nu; \alpha \beta} u_s(p).
\end{align}
The projection in terms of $\textbf B$ is then simply
\begin{align}
(\bar u(p) \Gamma u(p'),M) &= \text{tr} \left [ \Gamma_{\mu \nu; \alpha \beta} \textbf M^{\mu \nu; \alpha \beta} \right ],
\end{align}
where we have defined 
\begin{align}
\textbf M^{\mu \nu; \alpha \beta} = (\slashed p' - m) \textbf B^{\mu \nu; \alpha \beta} (\slashed p - m). 
\end{align}
In the following, we also use the definitions
\begin{align}
\widetilde{\textbf{M}}^{\mu \nu; \alpha \beta} = \frac{1}{2} \varepsilon^{\alpha \beta}_{~~~\lambda \kappa} \textbf{M}^{\mu \nu; \lambda \kappa}, \qquad \widetilde \sigma^{\mu \nu} = \frac{1}{2} \varepsilon^{\mu \nu}_{~~~\lambda \kappa} \sigma^{\lambda \kappa}.
\end{align}
Using the Gram-Schmidt procedure we find a total of 25 linear relations, leading to basis of 77 linearly independent tensors. We take for our subsequent analysis the particular choice of basis presented in  Appendix Eq.~\eqref{eq:tensornuc}. It is important to remark that this basis is not unique, since it depends on the order in which the initial set of 102 tensors is fed into the Gram-Schmidt algorithm. While in the spin-$0$ case the choice was essentially eliminating one of the 17 tensors appearing in the single identity in Eq.\,\eqref{eq: linrel}, for the spin-$\frac{1}{2}$ one has 25 such identities. 

After such a linearly independent basis is obtained it is straightforward to solve for the projections that give the following GITDs for $z^{\mu} = z^- n^{\mu}$:
\begin{align} \label{eq: Hg}
\frak H_g(\omega) &= \frac{1}{4(P^+)^3} \frac{1}{1-\xi^2} \text{tr} \left [ \left ( \gamma^+  + 2\xi^2 \frac{m}{\Delta_{\perp}^2}i\sigma^{+\Delta_{\perp}} \right ) \textbf M^{+i;+i} \right ],
\\
\frak E_g(\omega) &= \frac{1}{2(P^+)^3} \frac{m}{\Delta_{\perp}^2} \text{tr}\left [ i\sigma^{+\Delta_{\perp}} \textbf M^{+i;+i} \right ],
\\
\widetilde{\frak H}_g(\omega) &= \frac{1}{4(P^+)^3} \frac{1}{1-\xi^2}\text{tr} \left [ \left ( - i\gamma^+\gamma_5 - 2\xi \frac{m}{\Delta_{\perp}^2}i\widetilde \sigma^{+\Delta_{\perp}} \right ) \widetilde{\textbf M}^{+i;+i} \right ],
\\ \label{eq: tilde E}
\widetilde{\frak E}_g(\omega) &= \frac{1}{2(P^+)^3} \left ( - \frac{1}{\xi} \right ) \frac{m}{\Delta_{\perp}^2} \text{tr} \left [  i\widetilde{\sigma}^{+\Delta_{\perp}} \widetilde{\textbf M}^{+i;+i} \right ],
\\
\frak H_g^T(\omega) &= \frac{1}{2(P^+)^3} \frac{1}{1-\xi^2} \frac{m}{\Delta_{\perp}^2} \text{tr}\left [ \left ( 2m \xi \, i\gamma^+ \gamma_5 +  i\widetilde{\sigma}^{+\Delta_{\perp}} \right ) \left ( \widetilde{\textbf M}^{+1;+1} - \widetilde{\textbf M}^{+2;+2}\right )\right ],
\\ \notag
\frak E_g^T(\omega) &= \frac{1}{(P^+)^3} \frac{1}{1-\xi^2} \frac{m^2}{\Delta_{\perp}^2} \text{tr} \Bigg [ \left ( \gamma^+ + 2 \frac{m}{\Delta_{\perp}^2} i\sigma^{+\Delta_{\perp}} \right ) \left ( \textbf M^{+1;+1} - \textbf{M}^{+2;+2} \right ) 
\\
&\qquad - \left ( \xi\, i\gamma^+ \gamma_5 + 2 \frac{m}{\Delta_{\perp}^2} i \widetilde \sigma^{+\Delta_{\perp}} \right ) \left ( \widetilde{\textbf M}^{+1;+1} - \widetilde{\textbf M}^{+2;+2} \right ) \Bigg ],
\\
\widetilde{\frak H}_g^T(\omega) &= \frac{1}{(P^+)^3} \frac{m^3}{(\Delta_{\perp}^2)^2} \text{tr}\left [ i\sigma^{+\Delta_{\perp}} \left ( \textbf M^{+1;+1} - \textbf M^{+2;+2} \right ) - i\widetilde {\sigma}^{+\Delta_{\perp}} \left ( \widetilde{\textbf M}^{+1;+1} - \widetilde{\textbf M}^{+2;+2} \right ) \right ],
\\ \notag
\widetilde{\frak E}_g^T(\omega) &= \frac{1}{(P^+)^3} \frac{1}{1-\xi^2} \frac{m^2}{\Delta_{\perp}^2} \text{tr}\Bigg [ \xi \left ( \gamma^+ - 2 \frac{m}{\Delta_{\perp}^2} i\sigma^{+\Delta_{\perp}} \right ) \left ( \textbf M^{+1;+1} - \textbf M^{+2;+2} \right )
\\
&\quad - \left ( i\gamma^+ \gamma_5 - 2\xi \frac{m}{\Delta_{\perp}^2} i \widetilde{\sigma}^{+ \Delta_{\perp}} \right )  \left ( \widetilde{\textbf M}^{+1;+1} - \widetilde{\textbf M}^{+2;+2} \right ) \Bigg ], \label{eq: tildeEgT}
\end{align}
where
\begin{align}
\{ \frak H_g, \frak E_g, \wt{\frak H}_g, \wt{\frak E}_g, \frak H_g^T , \frak E_g^T, \wt {\frak H}_g^T, \wt {\frak E}_g^T \}(\omega) = \int_{-\infty}^{\infty} dx\,e^{ix\omega} \{  H_g,  E_g, \wt{ H}_g, \wt{ E}_g,  H_g^T ,  E_g^T, \wt { H}_g^T, \wt { E}_g^T \}(x).
\end{align}
Note that for $z^{\mu} = - z_3 \, \hat z^{\mu}$, one obtains a projection that matches onto the corresponding GITDs at the leading power by simply replacing the $+$'s in Eqs.\,\eqref{eq: Hg}-\eqref{eq: tildeEgT} by either $0$'s or $3$'s. To see this, note that for any  Lorentz four-vector $V^{\mu}$, which can be $P^{\mu}, \Delta^{\mu}$ or $\bar u(p')\gamma^{\mu} (\gamma_5) u(p),\bar u(p')\sigma^{\mu \ast} u(p)$ etc., we have
\begin{align}
V^{+} = V_0 + \mathcal O(1/P_3) = V_3 + \mathcal O(1/P_3).
\label{eq: V limit}
\end{align}
We reiterate that the $\mathcal O$ estimate is to be understood in the limit where $P_3 \rightarrow \infty$ at fixed $\omega, \eta, t, m^2$. Eq.\,\eqref{eq: V limit} follows from the behavior of $V^{\mu}$ under boosts in the $3$ direction, that is $V^{\pm} \rightarrow e^{\pm y} V^{\pm}$ for a boost with rapidity $y$. Note that Eq.\,\eqref{eq: V limit} also applies to $z^{\mu}$, since the $P_3 \rightarrow \infty$ at fixed $\omega$ implies $z_3 \rightarrow 0$. Hence changing the $+$'s in Eqs.\,\eqref{eq: Hg}-\eqref{eq: tildeEgT} by either $0$'s or $3$'s introduces additional terms of order $\mathcal O(1/P_3)$, which vanish in the infinite momentum limit.

However, it would not correspond to the exact linear combination $\sum_{\ell} f_{\ell} \mathcal M_{\ell}$ of invariant amplitudes such that
\begin{align}
\frak H_g = \sum_{\ell} f_{\ell} \mathcal M_{\ell} |_{z^2 = 0}.
\end{align}
The exact projections for our basis are presented in Appendix~\ref{sec:app}.

\section{Conclusion and Outlook}

We have worked out a Lorentz decomposition of the off-forward gluon matrix elements. This is nontrivial due to the linear dependence between Lorentz structures, which already appeared for the spin-$0$ case, see Sec.\,\ref{sec: spin0}. In fact, such linear dependencies also exist in the forward case for the polarized spin-$\frac{1}{2}$ hadron, see Appendix~\ref{app: forward}. The full decomposition for the spin-$\frac{1}{2}$ case was achieved using the Gram-Schmidt procedure to systematically determine linear relations. Using this basis we can systematically study the corresponding invariant amplitudes that are related to the GITDs.

As this work lays the foundation for performing LQCD calculations of various gluon GPDs, our next step is to carry out the perturbative matching within the LaMET and short-distance factorization based approaches~\cite{Yao:2022vtp} for the specific combinations of the off-forward matrix elements determined in this work. For the renormalization of the LQCD matrix elements, we will follow the standard procedures as the ratio~\cite{Radyushkin:2017cyf} and hybrid renormalization~\cite{Ji:2020brr} schemes. In addition to the challenge of achieving a good signal-to-noise ratio for the off-forward gluonic matrix elements, we note that reconstructing the $x$-dependence of GPDs from LQCD data when dealing with a finite set of discrete data can be challenging and state-of-the-art numerical techniques will be investigated. Reconstructing gluon GPDs is expected to be even more difficult due to their additional dependencies on $\xi$ and $t$, compared to the gluon PDFs~\cite{Chowdhury:2024ymm}. Additionally, the double distribution approach~\cite{Radyushkin:1998es,Radyushkin:1998bz,Radyushkin:2023ref} for extracting GPDs can also be explored. To extract the GPDs, one can also consider exploring various phenomenological models that provide a unified description of PDFs and GPDs~\cite{deTeramond:2021lxc}, as well as gluon GPDs at nonzero $\xi$~\cite{Chakrabarti:2024hwx}.

Finally, in this work, we have determined the off-forward gluonic matrix elements necessary for extracting all eight unpolarized, polarized, and transversity gluon GPDs in a lattice QCD calculation.  Notably, these projections have not been previously performed, and they will be essential for the numerical determination of gluon GPDs in future LQCD calculations. While gluon GPDs play a crucial role in our understanding of the spin and mass structures of the nucleon and other hadrons and their various emerging properties, due to the complexity of determining  gluon GPDs from the experiments, knowledge of the various gluon GPDs is not well established and some of them are virtually unknown. Lattice QCD calculations of gluon GPDs have the potential to open an avenue for elucidating the role of gluons in the hadrons.

\section{Acknowledgement}
We are grateful to Vladimir Braun and Alexander Manashov for their valuable suggestions and discussions, which significantly contributed to this research. We also extend our thanks to Anatoly Radyushkin for his insightful discussions and recommendations. We  acknowledge Shohini Bhattacharya, Martha Constantinou, Yao Ji, Huey-Wen Lin, Andreas Metz, and Swagato Mukherjee for stimulating discussions. J.S. is supported by the U.S. Department of Energy through Contract No. DE-SC0012704 and by Laboratory Directed Research and Development (LDRD) funds from Brookhaven Science Associates. R.S.S. is supported by Laboratory Directed Research and Development (LDRD No. 23-051) of BNL and RIKEN-BNL
Research Center. T.I. is supported by the U.S. Department of Energy (DOE) under Award No. DE-SC0012704, SciDAC-5 LAB 22-2580, and also Laboratory Directed Research and Development (LDRD No. 23-051) of BNL and RIKEN-BNL Research Center. Y.Y. is  supported in part by NSFC 
Grants No. 12293060, No. 12293062, and No. 12047503.

\section*{Data availability}
No data were created or analyzed in this study. 
\appendix

\section{Gordon identities}
\label{app: gordon}

In order to reduce the possible spinor structures $\bar u(p') \{ 1, \gamma_5 , \gamma^{\mu}, \gamma^{\mu} \gamma_5 , \sigma^{\mu \nu} \} u(p)$ that can appear, we use the following Gordon identities
\begin{align} \label{eq: G1}
\bar u(p') \gamma^{\mu} u(p) &= \frac{P^{\mu}}{m} \bar u(p') u(p) + \frac{i}{2m} \bar u(p') \sigma^{\mu \Delta} u(p) ,
\\ \label{eq: G2}
\bar u(p') i \sigma^{\mu P} u(p) &= - \frac{1}{2} \Delta^{\mu} \bar u(p') u(p),
\\ \label{eq: G3}
\bar u(p') \gamma^{\mu} \gamma_5 u(p) &= \frac{\Delta^{\mu}}{2m} \bar u(p') \gamma_5 u(p) - \frac{1}{2m} \varepsilon^{\mu P \nu \lambda} \bar u(p') \sigma_{\nu\lambda} u(p),
\\ \label{eq: G4}
\varepsilon^{\mu \Delta \alpha \beta} \bar u(p') i\sigma_{\alpha \beta} u(p) &= 4iP^{\mu} \bar u(p') \gamma_5 u(p),
\end{align}
which can be readily derived by inserting considering expressions $\bar u(p') (m \gamma^{\mu}(\gamma_5) \pm \gamma^{\mu}(\gamma_5) m ) u(p)$ and using the equations of motion $\slashed p u(p) = m u(p),\, \bar u(p') \slashed p' = m \bar u(p')$. Immediately, by the virtue of Eqs.\,\eqref{eq: G1} and \eqref{eq: G3}, we can eliminate $\gamma^{\mu}$ and $\gamma^{\mu} \gamma_5$ from the basis. Eq.\,\eqref{eq: G2} allows us to eliminate any structure where a $\sigma$ matrix is contracted with $P$ in favor of the scalar structure $\bar u(p') u(p)$. Finally, we can eliminate the pseudoscalar structure by the virtue of Eq.\,\eqref{eq: G4}. To see this, we multiply Eq.\,\eqref{eq: G4} with $P_{\mu} \varepsilon^{\mu_1 \mu_2 \mu_3 \mu_4}$, giving
\begin{align}
\varepsilon^{\mu_1 \mu_2 \mu_3 \mu_4} \bar u(p') \gamma_5 u(p) = \frac{1}{4P^2} \varepsilon^{\mu_1 \mu_2 \mu_3 \mu_4} \varepsilon^{P \Delta \alpha \beta} \bar u(p') \sigma_{\alpha \beta} u(p).
\end{align}
Rewriting the product of two Levi-Civita tensors on the right-hand-side in terms metric tensors, we obtain that $\varepsilon^{\mu_1 \mu_2 \mu_3 \mu_4}\bar u(p') \gamma_5 u(p)$, for arbitrary indices $\mu_j$, can be written in terms of the structure $\bar u(p') \sigma^{\mu \nu} u(p)$. Since parity conservation implies that $\bar u(p') \gamma_5 u(p)$ always has to appear with a Levi-Civita tensor and vice versa, we conclude that we can fully parametrize the spin structure of the matrix element by $\bar u(p') \{ 1, \sigma^{\mu \nu} \} u(p)$.

\section{Results for the spin-$\frac{1}{2}$ case}\label{sec:app}

In the following we use the shorthand notation
\begin{align}
S = \bar u(p') u(p), \qquad T^{\mu \nu} = \bar u(p') i \sigma^{\mu \nu} u(p).
\end{align}
The following list of tensors constitutes a (nonovercomplete) basis for the decomposition in Eq.\,\eqref{eq: Ansatz spinhalf} for $z^2 \leq 0$ and $\Delta_{\perp} \neq 0$ and $\xi \neq 0$.
\begin{align} \notag
 \mathcal{T}_1 &=S P^{\alpha } P^{\mu } g^{\beta  \nu }-S P^{\beta } P^{\mu } g^{\alpha  \nu }-S P^{\alpha } P^{\nu } g^{\beta  \mu }+S P^{\beta } P^{\nu } g^{\alpha  \mu } \\ \notag
 \mathcal{T}_2&=P^{\beta } P^{\nu } T^{\alpha  \mu }-P^{\alpha } P^{\nu } T^{\beta  \mu }-P^{\beta } P^{\mu } T^{\alpha  \nu }+P^{\alpha } P^{\mu } T^{\beta  \nu } \\ \notag
 \mathcal{T}_3 &=\Delta ^{\alpha } \Delta ^{\mu } T^{\beta  \nu }-\Delta ^{\beta } \Delta ^{\mu } T^{\alpha  \nu }-\Delta ^{\alpha } \Delta ^{\nu } T^{\beta  \mu }+\Delta ^{\beta } \Delta ^{\nu } T^{\alpha  \mu } \\ \notag
 \mathcal{T}_4 &=P^{\nu } \Delta ^{\alpha } \Delta ^{\mu } T^{\beta  \Delta }-P^{\nu } \Delta ^{\beta } \Delta ^{\mu } T^{\alpha  \Delta }-P^{\mu } \Delta ^{\alpha } \Delta ^{\nu } T^{\beta  \Delta }+P^{\mu } \Delta ^{\beta } \Delta ^{\nu } T^{\alpha  \Delta } \\ \notag
 \mathcal{T}_5 &=P^{\beta } \Delta ^{\alpha } \Delta ^{\mu } T^{\Delta  \nu }-P^{\alpha } \Delta ^{\beta } \Delta ^{\mu } T^{\Delta  \nu }-P^{\beta } \Delta ^{\alpha } \Delta ^{\nu } T^{\Delta  \mu }+P^{\alpha } \Delta ^{\beta } \Delta ^{\nu } T^{\Delta  \mu } \\ \notag
 \mathcal{T}_6 &=P^{\beta } P^{\nu } \Delta ^{\alpha } \Delta ^{\mu } T^{z \Delta }-P^{\alpha } P^{\nu } \Delta ^{\beta } \Delta ^{\mu } T^{z \Delta }-P^{\beta } P^{\mu } \Delta ^{\alpha } \Delta ^{\nu } T^{z \Delta }+P^{\alpha } P^{\mu } \Delta ^{\beta } \Delta ^{\nu } T^{z \Delta } \\ \notag
 \mathcal{T}_7 &=\Delta ^{\alpha } \Delta ^{\mu } g^{\beta  \nu } T^{z \Delta }-\Delta ^{\beta } \Delta ^{\mu } g^{\alpha  \nu } T^{z \Delta }-\Delta ^{\alpha } \Delta ^{\nu } g^{\beta  \mu } T^{z \Delta }+\Delta ^{\beta } \Delta ^{\nu } g^{\alpha  \mu } T^{z \Delta } \\ \notag
 \mathcal{T}_8 &=P^{\nu } \Delta ^{\alpha } \Delta ^{\mu } T^{z \beta }-P^{\nu } \Delta ^{\beta } \Delta ^{\mu } T^{z \alpha }-P^{\mu } \Delta ^{\alpha } \Delta ^{\nu } T^{z \beta }+P^{\mu } \Delta ^{\beta } \Delta ^{\nu } T^{z \alpha } \\ \notag
 \mathcal{T}_9 &=S P^{\beta } P^{\nu } \Delta ^{\alpha } \Delta ^{\mu }-S P^{\alpha } P^{\nu } \Delta ^{\beta } \Delta ^{\mu }-S P^{\beta } P^{\mu } \Delta ^{\alpha } \Delta ^{\nu }+S P^{\alpha } P^{\mu } \Delta ^{\beta } \Delta ^{\nu } \\ \notag
 \mathcal{T}_{10} &=S \Delta ^{\alpha } \Delta ^{\mu } g^{\beta  \nu }-S \Delta ^{\beta } \Delta ^{\mu } g^{\alpha  \nu }-S \Delta ^{\alpha } \Delta ^{\nu } g^{\beta  \mu }+S \Delta ^{\beta } \Delta ^{\nu } g^{\alpha  \mu } \\ \notag
 \mathcal{T}_{11}&=-P^{\nu } \Delta ^{\alpha } T^{\beta  \mu }+P^{\mu } \Delta ^{\alpha } T^{\beta  \nu }+P^{\nu } \Delta ^{\beta } T^{\alpha  \mu }-P^{\mu } \Delta ^{\beta } T^{\alpha  \nu } \\ \notag
 \mathcal{T}_{12}&=-P^{\beta } P^{\nu } \Delta ^{\alpha } T^{\Delta  \mu }+P^{\beta } P^{\mu } \Delta ^{\alpha } T^{\Delta  \nu }+P^{\alpha } P^{\nu } \Delta ^{\beta } T^{\Delta  \mu }-P^{\alpha } P^{\mu } \Delta ^{\beta } T^{\Delta  \nu } \\ \notag
 \mathcal{T}_{13} &=\Delta ^{\alpha } g^{\beta  \nu } T^{\Delta  \mu }-\Delta ^{\alpha } g^{\beta  \mu } T^{\Delta  \nu }-\Delta ^{\beta } g^{\alpha  \nu } T^{\Delta  \mu }+\Delta ^{\beta } g^{\alpha  \mu } T^{\Delta  \nu } \\ \notag
 \mathcal{T}_{14} &=P^{\mu } \Delta ^{\alpha } g^{\beta  \nu } T^{z \Delta }-P^{\nu } \Delta ^{\alpha } g^{\beta  \mu } T^{z \Delta }-P^{\mu } \Delta ^{\beta } g^{\alpha  \nu } T^{z \Delta }+P^{\nu } \Delta ^{\beta } g^{\alpha  \mu } T^{z \Delta } \\ \notag
 \mathcal{T}_{15} &=S P^{\mu } \Delta ^{\alpha } g^{\beta  \nu }-S P^{\nu } \Delta ^{\alpha } g^{\beta  \mu }-S P^{\mu } \Delta ^{\beta } g^{\alpha  \nu }+S P^{\nu } \Delta ^{\beta } g^{\alpha  \mu } \\ \notag
 \mathcal{T}_{16} &=P^{\nu } \Delta ^{\mu } T^{\alpha  \beta }-P^{\mu } \Delta ^{\nu } T^{\alpha  \beta } \\ \notag
 \mathcal{T}_{17} &=-P^{\beta } P^{\nu } \Delta ^{\mu } T^{\alpha  \Delta }+P^{\alpha } P^{\nu } \Delta ^{\mu } T^{\beta  \Delta }+P^{\beta } P^{\mu } \Delta ^{\nu } T^{\alpha  \Delta }-P^{\alpha } P^{\mu } \Delta ^{\nu } T^{\beta  \Delta } \\ \notag
 \mathcal{T}_{18} &=\Delta ^{\mu } g^{\beta  \nu } T^{\alpha  \Delta }-\Delta ^{\mu } g^{\alpha  \nu } T^{\beta  \Delta }-\Delta ^{\nu } g^{\beta  \mu } T^{\alpha  \Delta }+\Delta ^{\nu } g^{\alpha  \mu } T^{\beta  \Delta } \\ \notag
 \mathcal{T}_{19} &=P^{\alpha } \Delta ^{\mu } g^{\beta  \nu } T^{z \Delta }-P^{\beta } \Delta ^{\mu } g^{\alpha  \nu } T^{z \Delta }-P^{\alpha } \Delta ^{\nu } g^{\beta  \mu } T^{z \Delta }+P^{\beta } \Delta ^{\nu } g^{\alpha  \mu } T^{z \Delta } \\ \notag
 \mathcal{T}_{20}& =S P^{\alpha } \Delta ^{\mu } g^{\beta  \nu }-S P^{\beta } \Delta ^{\mu } g^{\alpha  \nu }-S P^{\alpha } \Delta ^{\nu } g^{\beta  \mu }+S P^{\beta } \Delta ^{\nu } g^{\alpha  \mu } \\ \notag
 \mathcal{T}_{21} &=P^{\mu } g^{\beta  \nu } T^{\alpha  \Delta }-P^{\nu } g^{\beta  \mu } T^{\alpha  \Delta }-P^{\mu } g^{\alpha  \nu } T^{\beta  \Delta }+P^{\nu } g^{\alpha  \mu } T^{\beta  \Delta } \\ \notag
 \mathcal{T}_{22}& =P^{\alpha } P^{\mu } g^{\beta  \nu } T^{z \Delta }-P^{\beta } P^{\mu } g^{\alpha  \nu } T^{z \Delta }-P^{\alpha } P^{\nu } g^{\beta  \mu } T^{z \Delta }+P^{\beta } P^{\nu } g^{\alpha  \mu } T^{z \Delta } \\ \notag
 \mathcal{T}_{23} &=S P^{\alpha } z^{\mu } g^{\beta  \nu }-S P^{\beta } z^{\mu } g^{\alpha  \nu }-S P^{\alpha } z^{\nu } g^{\beta  \mu }+S P^{\beta } z^{\nu } g^{\alpha  \mu } \\ \notag
 \mathcal{T}_{24} &=P^{\alpha } z^{\mu } g^{\beta  \nu } T^{z \Delta }-P^{\beta } z^{\mu } g^{\alpha  \nu } T^{z \Delta }-P^{\alpha } z^{\nu } g^{\beta  \mu } T^{z \Delta }+P^{\beta } z^{\nu } g^{\alpha  \mu } T^{z \Delta } \\ \notag
 \mathcal{T}_{25}& =z^{\mu } g^{\beta  \nu } T^{\alpha  \Delta }-z^{\mu } g^{\alpha  \nu } T^{\beta  \Delta }-z^{\nu } g^{\beta  \mu } T^{\alpha  \Delta }+z^{\nu } g^{\alpha  \mu } T^{\beta  \Delta } \\ \notag
 \mathcal{T}_{26} &=-P^{\beta } P^{\nu } z^{\mu } T^{\alpha  \Delta }+P^{\alpha } P^{\nu } z^{\mu } T^{\beta  \Delta }+P^{\beta } P^{\mu } z^{\nu } T^{\alpha  \Delta }-P^{\alpha } P^{\mu } z^{\nu } T^{\beta  \Delta } \\ \notag
 \mathcal{T}_{27} &=S P^{\mu } z^{\alpha } g^{\beta  \nu }-S P^{\nu } z^{\alpha } g^{\beta  \mu }-S P^{\mu } z^{\beta } g^{\alpha  \nu }+S P^{\nu } z^{\beta } g^{\alpha  \mu } \\ \notag
 \mathcal{T}_{28} &=P^{\mu } z^{\alpha } g^{\beta  \nu } T^{z \Delta }-P^{\nu } z^{\alpha } g^{\beta  \mu } T^{z \Delta }-P^{\mu } z^{\beta } g^{\alpha  \nu } T^{z \Delta }+P^{\nu } z^{\beta } g^{\alpha  \mu } T^{z \Delta } \\ \notag
 \mathcal{T}_{29}&=z^{\alpha } g^{\beta  \nu } T^{\Delta  \mu }-z^{\alpha } g^{\beta  \mu } T^{\Delta  \nu }-z^{\beta } g^{\alpha  \nu } T^{\Delta  \mu }+z^{\beta } g^{\alpha  \mu } T^{\Delta  \nu } \\ \notag
 \mathcal{T}_{30} &=-P^{\beta } P^{\nu } z^{\alpha } T^{\Delta  \mu }+P^{\beta } P^{\mu } z^{\alpha } T^{\Delta  \nu }+P^{\alpha } P^{\nu } z^{\beta } T^{\Delta  \mu }-P^{\alpha } P^{\mu } z^{\beta } T^{\Delta  \nu } \\ \notag
 \mathcal{T}_{31}&=-P^{\nu } z^{\alpha } T^{\beta  \mu }+P^{\mu } z^{\alpha } T^{\beta  \nu }+P^{\nu } z^{\beta } T^{\alpha  \mu }-P^{\mu } z^{\beta } T^{\alpha  \nu } \\ \notag
 \mathcal{T}_{32}&=S z^{\alpha } z^{\mu } g^{\beta  \nu }-S z^{\beta } z^{\mu } g^{\alpha  \nu }-S z^{\alpha } z^{\nu } g^{\beta  \mu }+S z^{\beta } z^{\nu } g^{\alpha  \mu } \\ \notag
 \mathcal{T}_{33} &=S P^{\beta } P^{\nu } z^{\alpha } z^{\mu }-S P^{\alpha } P^{\nu } z^{\beta } z^{\mu }-S P^{\beta } P^{\mu } z^{\alpha } z^{\nu }+S P^{\alpha } P^{\mu } z^{\beta } z^{\nu } \\ \notag
 \mathcal{T}_{34}& =P^{\beta } z^{\alpha } z^{\mu } T^{z \nu }-P^{\alpha } z^{\beta } z^{\mu } T^{z \nu }-P^{\beta } z^{\alpha } z^{\nu } T^{z \mu }+P^{\alpha } z^{\beta } z^{\nu } T^{z \mu } \\ \notag
 \mathcal{T}_{35}& =P^{\nu } z^{\alpha } z^{\mu } T^{z \beta }-P^{\nu } z^{\beta } z^{\mu } T^{z \alpha }-P^{\mu } z^{\alpha } z^{\nu } T^{z \beta }+P^{\mu } z^{\beta } z^{\nu } T^{z \alpha } \\ \notag
 \mathcal{T}_{36}& =z^{\alpha } z^{\mu } g^{\beta  \nu } T^{z \Delta }-z^{\beta } z^{\mu } g^{\alpha  \nu } T^{z \Delta }-z^{\alpha } z^{\nu } g^{\beta  \mu } T^{z \Delta }+z^{\beta } z^{\nu } g^{\alpha  \mu } T^{z \Delta } \\ \notag
 \mathcal{T}_{37} & =P^{\beta } P^{\nu } z^{\alpha } z^{\mu } T^{z \Delta }-P^{\alpha } P^{\nu } z^{\beta } z^{\mu } T^{z \Delta }-P^{\beta } P^{\mu } z^{\alpha } z^{\nu } T^{z \Delta }+P^{\alpha } P^{\mu } z^{\beta } z^{\nu } T^{z \Delta } \\ \notag
 \mathcal{T}_{38} &=P^{\beta } z^{\alpha } z^{\mu } T^{\Delta  \nu }-P^{\alpha } z^{\beta } z^{\mu } T^{\Delta  \nu }-P^{\beta } z^{\alpha } z^{\nu } T^{\Delta  \mu }+P^{\alpha } z^{\beta } z^{\nu } T^{\Delta  \mu } \\ \notag
 \mathcal{T}_{39} & =P^{\nu } z^{\alpha } z^{\mu } T^{\beta  \Delta }-P^{\nu } z^{\beta } z^{\mu } T^{\alpha  \Delta }-P^{\mu } z^{\alpha } z^{\nu } T^{\beta  \Delta }+P^{\mu } z^{\beta } z^{\nu } T^{\alpha  \Delta } \\ \notag
 \end{align}
 \begin{align} \notag
 \mathcal{T}_{40} &=z^{\alpha } z^{\mu } T^{\beta  \nu }-z^{\beta } z^{\mu } T^{\alpha  \nu }-z^{\alpha } z^{\nu } T^{\beta  \mu }+z^{\beta } z^{\nu } T^{\alpha  \mu } \\ \notag
 \mathcal{T}_{41} &=P^{\beta } \Delta ^{\mu } z^{\nu } T^{\alpha  \Delta }-P^{\alpha } \Delta ^{\mu } z^{\nu } T^{\beta  \Delta }-P^{\beta } \Delta ^{\nu } z^{\mu } T^{\alpha  \Delta }+P^{\alpha } \Delta ^{\nu } z^{\mu } T^{\beta  \Delta } \\ \notag
 \mathcal{T}_{42} &=\Delta ^{\nu } z^{\mu } T^{\alpha  \beta }-\Delta ^{\mu } z^{\nu } T^{\alpha  \beta } \\ \notag
 \mathcal{T}_{43} &=S \Delta ^{\mu } z^{\alpha } g^{\beta  \nu }-S \Delta ^{\mu } z^{\beta } g^{\alpha  \nu }-S \Delta ^{\nu } z^{\alpha } g^{\beta  \mu }+S \Delta ^{\nu } z^{\beta } g^{\alpha  \mu } \\ \notag
 \mathcal{T}_{44} &=S P^{\beta } P^{\nu } \Delta ^{\mu } z^{\alpha }-S P^{\alpha } P^{\nu } \Delta ^{\mu } z^{\beta }-S P^{\beta } P^{\mu } \Delta ^{\nu } z^{\alpha }+S P^{\alpha } P^{\mu } \Delta ^{\nu } z^{\beta } \\ \notag
 \mathcal{T}_{45} &=P^{\nu } \Delta ^{\mu } z^{\alpha } T^{z \beta }-P^{\nu } \Delta ^{\mu } z^{\beta } T^{z \alpha }-P^{\mu } \Delta ^{\nu } z^{\alpha } T^{z \beta }+P^{\mu } \Delta ^{\nu } z^{\beta } T^{z \alpha } \\ \notag
 \mathcal{T}_{46} &=\Delta ^{\mu } z^{\alpha } g^{\beta  \nu } T^{z \Delta }-\Delta ^{\mu } z^{\beta } g^{\alpha  \nu } T^{z \Delta }-\Delta ^{\nu } z^{\alpha } g^{\beta  \mu } T^{z \Delta }+\Delta ^{\nu } z^{\beta } g^{\alpha  \mu } T^{z \Delta } \\ \notag
 \mathcal{T}_{47} &=P^{\beta } P^{\nu } \Delta ^{\mu } z^{\alpha } T^{z \Delta }-P^{\alpha } P^{\nu } \Delta ^{\mu } z^{\beta } T^{z \Delta }-P^{\beta } P^{\mu } \Delta ^{\nu } z^{\alpha } T^{z \Delta }+P^{\alpha } P^{\mu } \Delta ^{\nu } z^{\beta } T^{z \Delta } \\ \notag
 \mathcal{T}_{48}& =P^{\beta } \Delta ^{\mu } z^{\alpha } T^{\Delta  \nu }-P^{\alpha } \Delta ^{\mu } z^{\beta } T^{\Delta  \nu }-P^{\beta } \Delta ^{\nu } z^{\alpha } T^{\Delta  \mu }+P^{\alpha } \Delta ^{\nu } z^{\beta } T^{\Delta  \mu } \\ \notag
 \mathcal{T}_{49} & =P^{\nu } \Delta ^{\mu } z^{\alpha } T^{\beta  \Delta }-P^{\nu } \Delta ^{\mu } z^{\beta } T^{\alpha  \Delta }-P^{\mu } \Delta ^{\nu } z^{\alpha } T^{\beta  \Delta }+P^{\mu } \Delta ^{\nu } z^{\beta } T^{\alpha  \Delta } \\ \notag
 \mathcal{T}_{50} &=\Delta ^{\mu } z^{\alpha } T^{\beta  \nu }-\Delta ^{\mu } z^{\beta } T^{\alpha  \nu }-\Delta ^{\nu } z^{\alpha } T^{\beta  \mu }+\Delta ^{\nu } z^{\beta } T^{\alpha  \mu } \\ \notag
 \mathcal{T}_{51} &=-S P^{\beta } \Delta ^{\mu } z^{\alpha } z^{\nu }+S P^{\alpha } \Delta ^{\mu } z^{\beta } z^{\nu }+S P^{\beta } \Delta ^{\nu } z^{\alpha } z^{\mu }-S P^{\alpha } \Delta ^{\nu } z^{\beta } z^{\mu } \\ \notag
 \mathcal{T}_{52} &=\Delta ^{\mu } z^{\alpha } z^{\nu } \left(-T^{z \beta }\right)+\Delta ^{\mu } z^{\beta } z^{\nu } T^{z \alpha }+\Delta ^{\nu } z^{\alpha } z^{\mu } T^{z \beta }-\Delta ^{\nu } z^{\beta } z^{\mu } T^{z \alpha } \\ \notag
 \mathcal{T}_{53} &=\Delta ^{\mu } z^{\alpha } z^{\nu } \left(-T^{\beta  \Delta }\right)+\Delta ^{\mu } z^{\beta } z^{\nu } T^{\alpha  \Delta }+\Delta ^{\nu } z^{\alpha } z^{\mu } T^{\beta  \Delta }-\Delta ^{\nu } z^{\beta } z^{\mu } T^{\alpha  \Delta } \\ \notag
 \mathcal{T}_{54} &=S \Delta ^{\alpha } z^{\mu } g^{\beta  \nu }-S \Delta ^{\alpha } z^{\nu } g^{\beta  \mu }-S \Delta ^{\beta } z^{\mu } g^{\alpha  \nu }+S \Delta ^{\beta } z^{\nu } g^{\alpha  \mu } \\ \notag
 \mathcal{T}_{55} &=S P^{\beta } P^{\nu } \Delta ^{\alpha } z^{\mu }-S P^{\beta } P^{\mu } \Delta ^{\alpha } z^{\nu }-S P^{\alpha } P^{\nu } \Delta ^{\beta } z^{\mu }+S P^{\alpha } P^{\mu } \Delta ^{\beta } z^{\nu } \\ \notag
 \mathcal{T}_{56} &=P^{\nu } \Delta ^{\alpha } z^{\mu } T^{z \beta }-P^{\mu } \Delta ^{\alpha } z^{\nu } T^{z \beta }-P^{\nu } \Delta ^{\beta } z^{\mu } T^{z \alpha }+P^{\mu } \Delta ^{\beta } z^{\nu } T^{z \alpha } \\ \notag
 \mathcal{T}_{57}& =\Delta ^{\alpha } z^{\mu } g^{\beta  \nu } T^{z \Delta }-\Delta ^{\alpha } z^{\nu } g^{\beta  \mu } T^{z \Delta }-\Delta ^{\beta } z^{\mu } g^{\alpha  \nu } T^{z \Delta }+\Delta ^{\beta } z^{\nu } g^{\alpha  \mu } T^{z \Delta } \\ \notag
 \mathcal{T}_{58} &=P^{\beta } P^{\nu } \Delta ^{\alpha } z^{\mu } T^{z \Delta }-P^{\beta } P^{\mu } \Delta ^{\alpha } z^{\nu } T^{z \Delta }-P^{\alpha } P^{\nu } \Delta ^{\beta } z^{\mu } T^{z \Delta }+P^{\alpha } P^{\mu } \Delta ^{\beta } z^{\nu } T^{z \Delta } \\ \notag
 \mathcal{T}_{59} &=P^{\beta } \Delta ^{\alpha } z^{\mu } T^{\Delta  \nu }-P^{\beta } \Delta ^{\alpha } z^{\nu } T^{\Delta  \mu }-P^{\alpha } \Delta ^{\beta } z^{\mu } T^{\Delta  \nu }+P^{\alpha } \Delta ^{\beta } z^{\nu } T^{\Delta  \mu } \\ \notag
 \mathcal{T}_{60} & =P^{\nu } \Delta ^{\alpha } z^{\mu } T^{\beta  \Delta }-P^{\mu } \Delta ^{\alpha } z^{\nu } T^{\beta  \Delta }-P^{\nu } \Delta ^{\beta } z^{\mu } T^{\alpha  \Delta }+P^{\mu } \Delta ^{\beta } z^{\nu } T^{\alpha  \Delta } \\ \notag
 \mathcal{T}_{61} &=\Delta ^{\alpha } z^{\mu } T^{\beta  \nu }-\Delta ^{\alpha } z^{\nu } T^{\beta  \mu }-\Delta ^{\beta } z^{\mu } T^{\alpha  \nu }+\Delta ^{\beta } z^{\nu } T^{\alpha  \mu } \\ \notag
 \mathcal{T}_{62} &=P^{\nu } \Delta ^{\alpha } z^{\beta } T^{\Delta  \mu }-P^{\mu } \Delta ^{\alpha } z^{\beta } T^{\Delta  \nu }-P^{\nu } \Delta ^{\beta } z^{\alpha } T^{\Delta  \mu }+P^{\mu } \Delta ^{\beta } z^{\alpha } T^{\Delta  \nu } \\ \notag
 \mathcal{T}_{63}& =\Delta ^{\beta } z^{\alpha } T^{\mu  \nu }-\Delta ^{\alpha } z^{\beta } T^{\mu  \nu } \\ \notag
 \mathcal{T}_{64}&=-S P^{\nu } \Delta ^{\alpha } z^{\beta } z^{\mu }+S P^{\mu } \Delta ^{\alpha } z^{\beta } z^{\nu }+S P^{\nu } \Delta ^{\beta } z^{\alpha } z^{\mu }-S P^{\mu } \Delta ^{\beta } z^{\alpha } z^{\nu } \\ \notag
 \mathcal{T}_{65}&=\Delta ^{\alpha } z^{\beta } z^{\mu } \left(-T^{z \nu }\right)+\Delta ^{\alpha } z^{\beta } z^{\nu } T^{z \mu }+\Delta ^{\beta } z^{\alpha } z^{\mu } T^{z \nu }-\Delta ^{\beta } z^{\alpha } z^{\nu } T^{z \mu } \\ \notag
 \mathcal{T}_{66}& =-P^{\nu } \Delta ^{\alpha } z^{\beta } z^{\mu } T^{z \Delta }+P^{\mu } \Delta ^{\alpha } z^{\beta } z^{\nu } T^{z \Delta }+P^{\nu } \Delta ^{\beta } z^{\alpha } z^{\mu } T^{z \Delta }-P^{\mu } \Delta ^{\beta } z^{\alpha } z^{\nu } T^{z \Delta } \\ \notag
 \mathcal{T}_{67}& =\Delta ^{\alpha } z^{\beta } z^{\mu } \left(-T^{\Delta  \nu }\right)+\Delta ^{\alpha } z^{\beta } z^{\nu } T^{\Delta  \mu }+\Delta ^{\beta } z^{\alpha } z^{\mu } T^{\Delta  \nu }-\Delta ^{\beta } z^{\alpha } z^{\nu } T^{\Delta  \mu } \\ \notag
 \mathcal{T}_{68} &=-S P^{\beta } \Delta ^{\alpha } \Delta ^{\mu } z^{\nu }+S P^{\alpha } \Delta ^{\beta } \Delta ^{\mu } z^{\nu }+S P^{\beta } \Delta ^{\alpha } \Delta ^{\nu } z^{\mu }-S P^{\alpha } \Delta ^{\beta } \Delta ^{\nu } z^{\mu } \\ \notag
 \mathcal{T}_{69} &=\Delta ^{\alpha } \Delta ^{\mu } z^{\nu } \left(-T^{z \beta }\right)+\Delta ^{\beta } \Delta ^{\mu } z^{\nu } T^{z \alpha }+\Delta ^{\alpha } \Delta ^{\nu } z^{\mu } T^{z \beta }-\Delta ^{\beta } \Delta ^{\nu } z^{\mu } T^{z \alpha } \\ \notag
 \mathcal{T}_{70}&=-P^{\beta } \Delta ^{\alpha } \Delta ^{\mu } z^{\nu } T^{z \Delta }+P^{\alpha } \Delta ^{\beta } \Delta ^{\mu } z^{\nu } T^{z \Delta }+P^{\beta } \Delta ^{\alpha } \Delta ^{\nu } z^{\mu } T^{z \Delta }-P^{\alpha } \Delta ^{\beta } \Delta ^{\nu } z^{\mu } T^{z \Delta } \\ \notag
 \mathcal{T}_{71}&=\Delta ^{\alpha } \Delta ^{\mu } z^{\nu } \left(-T^{\beta  \Delta }\right)+\Delta ^{\beta } \Delta ^{\mu } z^{\nu } T^{\alpha  \Delta }+\Delta ^{\alpha } \Delta ^{\nu } z^{\mu } T^{\beta  \Delta }-\Delta ^{\beta } \Delta ^{\nu } z^{\mu } T^{\alpha  \Delta } \\ \notag
 \mathcal{T}_{72}& =-S P^{\nu } \Delta ^{\alpha } \Delta ^{\mu } z^{\beta }+S P^{\nu } \Delta ^{\beta } \Delta ^{\mu } z^{\alpha }+S P^{\mu } \Delta ^{\alpha } \Delta ^{\nu } z^{\beta }-S P^{\mu } \Delta ^{\beta } \Delta ^{\nu } z^{\alpha } \\ \notag
 \mathcal{T}_{73}&=\Delta ^{\alpha } \Delta ^{\mu } z^{\beta } \left(-T^{z \nu }\right)+\Delta ^{\beta } \Delta ^{\mu } z^{\alpha } T^{z \nu }+\Delta ^{\alpha } \Delta ^{\nu } z^{\beta } T^{z \mu }-\Delta ^{\beta } \Delta ^{\nu } z^{\alpha } T^{z \mu } \\ \notag
 \mathcal{T}_{74}&=-P^{\nu } \Delta ^{\alpha } \Delta ^{\mu } z^{\beta } T^{z \Delta }+P^{\nu } \Delta ^{\beta } \Delta ^{\mu } z^{\alpha } T^{z \Delta }+P^{\mu } \Delta ^{\alpha } \Delta ^{\nu } z^{\beta } T^{z \Delta }-P^{\mu } \Delta ^{\beta } \Delta ^{\nu } z^{\alpha } T^{z \Delta } \\ \notag
 \mathcal{T}_{75} &=\Delta ^{\alpha } \Delta ^{\mu } z^{\beta } \left(-T^{\Delta  \nu }\right)+\Delta ^{\beta } \Delta ^{\mu } z^{\alpha } T^{\Delta  \nu }+\Delta ^{\alpha } \Delta ^{\nu } z^{\beta } T^{\Delta  \mu }-\Delta ^{\beta } \Delta ^{\nu } z^{\alpha } T^{\Delta  \mu } \\ \notag
 \mathcal{T}_{76}& =S \Delta ^{\alpha } \Delta ^{\mu } z^{\beta } z^{\nu }-S \Delta ^{\beta } \Delta ^{\mu } z^{\alpha } z^{\nu }-S \Delta ^{\alpha } \Delta ^{\nu } z^{\beta } z^{\mu }+S \Delta ^{\beta } \Delta ^{\nu } z^{\alpha } z^{\mu } \\ 
 \mathcal{T}_{77} &=\Delta ^{\alpha } \Delta ^{\mu } z^{\beta } z^{\nu } T^{z \Delta }-\Delta ^{\beta } \Delta ^{\mu } z^{\alpha } z^{\nu } T^{z \Delta }-\Delta ^{\alpha } \Delta ^{\nu } z^{\beta } z^{\mu } T^{z \Delta }+\Delta ^{\beta } \Delta ^{\nu } z^{\alpha } z^{\mu } T^{z \Delta } \label{eq:tensornuc}
\end{align}
%%%%
The first 22 structures correspond to the invariant amplitudes $\mathcal M_1$ to $\mathcal M_{22}$ that contribute to the GITDs at $z^2 = 0$.
In terms of this basis the GITDs defined in Eqs.\,\eqref{eq: Hg}-\eqref{eq: tildeEgT} become ($z^2 = 0$ implied)
\begin{align} \notag
\frak H_g &=-2 \mathcal{M}_1 +2 \xi ^2 t \mathcal{M}_4 +2 \xi ^2 t \mathcal{M}_5 +4 \xi ^3 \omega  \mathcal{M}_8 -t \mathcal{M}_9 -4 m^2 \xi ^2 \mathcal{M}_9+\xi ^2 t \mathcal{M}_9 -8 \xi ^2 \mathcal{M}_{10} -2 \xi ^2 \mathcal{M}_{11} \\ \label{eq: Hg Ms}
    &\quad  -\xi  t \mathcal{M}_{12} +4 \xi  \mathcal{M}_{15}+2 \xi ^2 \mathcal{M}_{16}-\xi  t \mathcal{M}_{17}+4 \xi  \mathcal{M}_{20}, \\ \notag
\frak E_g &=2 \mathcal{M}_1+2 t \mathcal{M}_4 -2 \xi ^2 t \mathcal{M}_4 +2 t \mathcal{M}_5-2 \xi ^2 t \mathcal{M}_5 -2 t \omega  \mathcal{M}_6 -8 m^2 \xi ^2 \omega  \mathcal{M}_6 +2 \xi ^2 t \omega  \mathcal{M}_6 -16 \xi ^2 \omega  \mathcal{M}_7 \\ \notag
&\quad +4 \xi  \omega  \mathcal{M}_8-4 \xi ^3 \omega  \mathcal{M}_8 +t \mathcal{M}_9+4 m^2 \xi ^2 \mathcal{M}_9-\xi ^2 t \mathcal{M}_9 +8 \xi ^2 \mathcal{M}_{10}-2 \mathcal{M}_{11}+2 \xi ^2 \mathcal{M}_{11}+4 m^2 \xi  \mathcal{M}_{12} \\ \notag
&\quad -8 \xi  \mathcal{M}_{13}+8 \xi  \omega  \mathcal{M}_{14}-4 \xi  \mathcal{M}_{15}+2 \mathcal{M}_{16}-2 \xi ^2 \mathcal{M}_{16}+4 m^2 \xi  \mathcal{M}_{17}-8 \xi  \mathcal{M}_{18}+8 \xi  \omega  \mathcal{M}_{19}
-4 \xi  \mathcal{M}_{20}\\ 
&\quad  +4 \mathcal{M}_{21}-4 \omega  \mathcal{M}_{22},   \\ 
   \widetilde{\frak H}_g &=-2 \mathcal{M}_2-2 \xi  t \mathcal{M}_4+2 \xi  t \mathcal{M}_5-4 \xi ^2 \omega  \mathcal{M}_8+2 \xi  \mathcal{M}_{11}-t \mathcal{M}_{12}-2 \xi  \mathcal{M}_{16}+t \mathcal{M}_{17}, \\ \notag
   \widetilde{\frak E}_g &=2 \mathcal{M}_2+8 \mathcal{M}_3+8 m^2 \xi  \mathcal{M}_4-8 m^2 \xi  \mathcal{M}_5+4 \xi ^2 \omega  \mathcal{M}_8-4 \omega  \mathcal{M}_8 -2 \xi  \mathcal{M}_{11}-\frac{2 \mathcal{M}_{11}}{\xi }+4 m^2 \mathcal{M}_{12}\\ 
    &\quad  -\frac{2 \mathcal{M}_{16}}{\xi }+2 \xi  \mathcal{M}_{16}-4 m^2 \mathcal{M}_{17}, \\ 
\frak H_g^T&=4 \xi  \omega  \mathcal{M}_8-2 \mathcal{M}_{11}+2 \mathcal{M}_{16},    \\
\frak E_g^T&=8 m^2 \mathcal{M}_4+8 m^2 \mathcal{M}_5 -8 m^2 \omega  \mathcal{M}_6, \\
 \widetilde{\frak H}_g^T&=-4 m^2 \mathcal{M}_4-4 m^2 \mathcal{M}_5 + 4 m^2 \omega  \mathcal{M}_6-2 m^2 \mathcal{M}_9, \\
 \widetilde{\frak E}_g^T& =-8 m^2 \xi  \omega  \mathcal{M}_6+4 m^2 \mathcal{M}_{12}+4 m^2 \mathcal{M}_{17}. \label{eq: EgT Ms}
\end{align}

Note that some of the coefficients of the $\mathcal M_{\ell}$ are identical, the reason typically being that the corresponding tensors are related by permutation symmetry $(\mu , \nu, z ) \leftrightarrow (\alpha, \beta, -z)$.

In the following, we present our results for the projections that obey
\begin{align}
(\mathcal P[\frak F], M) = \sum_{\ell} f_{\ell}^{(\frak F)} \mathcal M_{\ell},
\end{align}
where $\frak F \in \{ \frak H_g, \frak E_g, \widetilde{\frak H}_g, \widetilde{\frak E}_g, \frak H_g^T, \frak E_g^T, \widetilde{\frak H}_g^T, \widetilde{\frak E}_g^T \}$ and the coefficients $f_{\ell}^{(\frak F)} = f_{\ell}^{(\frak F)}(\omega, \eta, t)$, that are defined by
\begin{align}
\frak F = \sum_{\ell} f_{\ell}^{(\frak F)} \mathcal M_{\ell} |_{z^2 = 0},
\end{align}
can be read of from Eqs.\,\eqref{eq: Hg Ms}--\eqref{eq: EgT Ms}. Note that $f_{11}^{(\widetilde{\frak E}_g)}$ and $f_{16}^{(\widetilde{\frak E}_g)}$ have a simple pole in $\xi$. This is of course due to the factor of $\frac{1}{\xi}$ in Eq.\,\eqref{eq: tilde E}. Such poles are not unexpected. Given that the $\widetilde{\frak E}_g$ is regular at $\xi = 0$, this just implies that $\text{tr} \left [  i\widetilde{\sigma}^{+\Delta_{\perp}} \widetilde{\textbf M}^{+i;+i} \right ]$ as well as $\mathcal M_{11}$, $\mathcal M_{16}$ vanish at least linearly as $\xi \rightarrow 0$. 

We now present our results for the projections\footnote{The expressions can be further rearranged by combining various matrix elements, for example ${\bf M}_{02;12}$ and ${\bf M}_{12;02}$, etc. but we still write them separately in the following.}.

\begin{align} \notag
{\rm Projection~onto}~\frak{H}_g:\\ \notag
(\mathcal{P}[\frak{H}_g],M) &=-\frac{2 \xi ^2 z_3^2 }{3 m\eta ^2  \omega ^2} \text{tr}[\textbf M_{01;01} + \textbf M_{02;02}] \\ \notag 
&+ \frac{\xi ^3 z_3  \left(\frac{12 \omega ^2 (\eta -\xi )}{\xi }+z_3^2 \left(t-4 m^2\right)\right)}{3 m\Delta _1 \eta ^2  \omega ^3} \text{tr}[{\bf M}_{02;12} + {\bf M}_{12;02}]\\ \notag 
& -\frac{\xi   \left(8 \xi ^2 \omega ^2+\Delta _1^2 z_3^2\right) \left(8 \omega ^2 (\xi -\eta )+\xi  z_3^2 \left(4 m^2-t\right)\right)}{6 \Delta _1^2 \eta ^2 m \omega ^4} \text{tr}[{\bf M}_{12;12}] \\ \notag
&-\frac{\xi ^4 z_3^2 \left(4 m^2-t\right)  \left(4 m^2 \xi  (\eta -\xi )+t\right)}{4 \Delta _1^2 \eta ^2 m t \omega ^2} \text{tr}[i \gamma _5{\bf M}_{01;02} + i \gamma _5{\bf M}_{02;01}]  \\ \notag
& +\frac{\xi ^2 z_3 (\eta -\xi ) \left(4 m^2 \xi ^2-t\right) }{2 \Delta _1 \eta ^2 m t \omega }  \text{tr}[i \gamma _5{\bf M}_{01;12}+i \gamma _5{\bf M}_{12;01}]   \\\ \notag 
& + \frac{\xi ^2  \left(21 \omega ^2 (\eta -\xi )+z_3^2 \left(m^2 (5 \eta -21 \xi )-2 t (\eta -3 \xi )\right)\right)}{3 \Delta _1^2 \eta ^2 m \omega ^2} \text{tr}[i \sigma_{10}{\bf M}_{02;12}+i \sigma_{10}{\bf M}_{12;02}]      \\ \notag
&-\frac{\xi ^2 t z_3^2 (\eta -\xi ) }{4 \Delta _1^2 \eta ^2 m \omega ^2}\text{tr}[i \sigma_{20}{\bf M}_{12;01}]\\ \notag
& + \frac{\xi ^3 t z_3^2 }{4 \eta ^2 m \omega ^2 \left(\xi ^2 \left(4 m^2-t\right)+t\right)}
\text{tr}[i \sigma_{21}{\bf M}_{01;02}+i \sigma_{21}{\bf M}_{02;01}]\\ \notag 
&-\frac{8 z_3^2 }{12 m^3 \xi  z_3^2+12 m \xi  \omega ^2-3 m \xi  t z_3^2}\text{tr}[i \sigma_{30}{\bf M}_{01;01}+i \sigma_{30}{\bf M}_{02;02}]  \\ \notag
&+\frac{\xi ^3  \left(\frac{84 \omega ^2 (\eta -\xi )}{\xi }+5 z_3^2 \left(t-4 m^2\right)\right)}{12 \eta ^2 m \omega ^2 \left(\xi ^2 \left(4 m^2-t\right)+t\right)}\text{tr}[i \sigma_{31}{\bf M}_{02;12} + i \sigma_{31}{\bf M}_{12;02}]  \\ \notag
&+ \frac{t z_3^3  \left(4 \omega  \left(\omega -\frac{2 \eta  \omega }{\xi }\right)+z_3^2 \left(4 m^2-t\right)\right)}{2 \Delta _1 m \omega  \left(z_3^2 \left(t-4 m^2\right)-4 \omega ^2\right){}^2} \text{tr}[i \sigma_{31}{\bf M}_{12;12}]\\ \notag
&+ \frac{\xi ^4 t z_3^2  \left(4 \omega ^2 (\xi -\eta )+\xi  z_3^2 \left(4 m^2-t\right)\right)}{16 \Delta _1^2 \eta ^4 m \omega ^4} \text{tr}[i \sigma_{32}{\bf M}_{01;12}]\\ \notag
&+ \frac{1}{{48 \Delta _1^2 \eta ^4 m \omega ^6}}\bigg(\xi ^2 \big(512 \xi ^2 \omega ^6 (\eta -\xi )-4 \omega ^4 z_3^2 \big(4 m^2 \xi ^2 (35 \xi -22 \eta )+t \big(22 \eta  \xi ^2+10 \eta -35 \xi ^3-13 \xi \big)\big)\\ \notag
&+\omega ^2 z_3^4 \big(-32 m^4 \xi ^2 (5 \eta -\xi )+4 m^2 t \big(26 \eta  \xi ^2-10 \eta -7 \xi ^3+18 \xi \big)+t^2 \big(-16 \eta  \xi ^2+16 \eta +5 \xi ^3-21 \xi \big)\big)\\ 
&+\xi  z_3^6 \big(20 m^4-13 m^2 t+2 t^2\big) \big(4 m^2 \xi ^2-\xi ^2 t+t\big)\big)\bigg) \text{tr}[i \sigma_{30}{\bf M}_{12;12}]
\end{align}
%%%%%
%%%%%%

\begin{align} \notag
{\rm Projection~onto}~\frak{E}_g:\\ \notag
(\mathcal{P}[\frak{E}_g],M) &= \frac{32 m z_3^2 }{12 t \omega ^2-3 t z_3^2 \left(t-4 m^2\right)}\text{tr}[{\bf M}_{01;01}+ {\bf M}_{02;02}]\\ \notag
&+\frac{4 m \xi ^2 z_3  \left(12 \omega ^2 (\xi -\eta )+\xi  z_3^2 \left(4 m^2-t\right)\right)}{3 \Delta _1 \eta ^2 t \omega ^3} \text{tr}[{\bf M}_{02;12}+{\bf M}_{12;02}] \\ \notag
&+ \frac{2 m \xi   \left(8 \xi ^2 \omega ^2+\Delta _1^2 z_3^2\right) \left(8 \omega ^2 (\xi -\eta )+\xi  z_3^2 \left(4 m^2-t\right)\right)}{3 \Delta _1^2 \eta ^2 t \omega ^4} \text{tr}[{\bf M}_{12;12}]\\ \notag
&-\frac{m \xi ^3 z_3^2 \left(-\eta  \xi ^2+\eta +\xi ^3-2 \xi \right) \left(4 m^2-t\right) }{\Delta _1^2 \eta ^2 t \omega ^2}\text{tr}[i \gamma _5{\bf M}_{01;02}+i \gamma _5{\bf M}_{02;01}]\\ \notag
&-\frac{2 m \xi ^2 \left(\xi ^2-2\right) z_3 (\eta -\xi ) }{\Delta _1 \eta ^2 t \omega } \text{tr}[i \gamma _5{\bf M}_{01;12} + i \gamma _5{\bf M}_{12;01}]\\ \notag
& +\frac{2 m \xi ^2  \left(30 \omega ^2 (\xi -\eta )+z_3^2 \left(t (4 \eta -9 \xi )-10 m^2 (\eta -3 \xi )\right)\right)}{3 \Delta _1^2 \eta ^2 t \omega ^2} \text{tr}[i \sigma_{10}{\bf M}_{02;12}+i \sigma_{10}{\bf M}_{12;02}]  \\ \notag
& + \frac{m \xi ^2 z_3^2 (\eta -\xi ) }{\Delta _1^2 \eta ^2 \omega ^2}\text{tr}[i \sigma_{20}{\bf M}_{12;01}]\\ \notag
&+ \frac{m \xi ^3 z_3^2 }{\Delta _1^2 \eta ^2 \omega ^2} \text{tr}[i \sigma_{21}{\bf M}_{01;02}+i \sigma_{21}{\bf M}_{02;01}] \\ \notag
&-\frac{20 m z_3^2 }{3 \xi  t \left(z_3^2 \left(t-4 m^2\right)-4 \omega ^2\right)} \text{tr}[i \sigma_{30}{\bf M}_{01;01})+i \sigma_{30}{\bf M}_{02;02}]\\ \notag
&+ \frac{m \xi   \left(80 \xi ^2 \omega ^4 (\eta -\xi )-z_3^4 (\eta -2 \xi ) \left(5 m^2-2 t\right) \left(4 m^2 \xi ^2-\xi ^2 t+t\right)-10 \omega ^2 z_3^2 \left(4 m^2 \xi ^3+t \left(\eta -\xi  \left(\xi ^2+1\right)\right)\right)\right)}{3 \Delta _1^2 \eta ^3 t \omega ^4}\\ \notag
&~\text{tr}[i \sigma_{30}{\bf M}_{12;12}]\\ \notag
& - \frac{5 m \xi ^2  \left(12 \omega ^2 (\xi -\eta )+\xi  z_3^2 \left(4 m^2-t\right)\right)}{3 \Delta _1^2 \eta ^2 t \omega ^2}\text{tr}[i \sigma_{31}{\bf M}_{02;12} + i \sigma_{31}{\bf M}_{12;02}] \\\notag
& + \frac{m \xi ^4 z_3^3  \left(4 \omega ^2 \left(\frac{2 \eta }{\xi }-1\right)+z_3^2 \left(t-4 m^2\right)\right)}{8 \Delta _1 \eta ^4 \omega ^5}\text{tr}[i \sigma_{31}{\bf M}_{12;12}]\\ 
&-\frac{m \xi ^3 z_3^2 (\eta -\xi ) }{\Delta _1^2 \eta ^3 \omega ^2}\text{tr}[i \sigma_{32}{\bf M}_{01;12}]
\end{align}

%%%%%
\begin{align} \notag
{\rm Projection~onto}~\wt{\frak{H}}_g:\\ \notag
 (\mathcal{P}[\wt{\frak{H}}_g],M) &=   \frac{\xi   \left(\xi  z_3^3 \left(4 m^2-t\right)+4 \omega ^2 z_3 (\xi -\eta )\right)}{3 \Delta _1 \eta ^2 m \omega ^3} \text{tr}[{\bf M}_{02;12}]\\ \notag
 &+\frac{\xi  z_3  \left(16 \xi ^2 \omega ^4 (\eta -\xi )+4 \omega ^2 z_3^2 \left(4 \eta  m^2 \xi ^2+t \left(-\eta  \xi ^2+\eta +\xi \right)\right)+\xi  z_3^4 \left(4 m^2-t\right) \left(4 m^2 \xi ^2-\xi ^2 t+t\right)\right)}{3 \Delta _1 \eta ^2 m \omega ^3 \left(4 \xi ^2 \omega ^2-\Delta _1^2 z_3^2\right)}\text{tr}[{\bf M}_{12;02}]\\ \notag
 & + \frac{2 z_3^2  \left(\frac{1}{\omega ^2}+\frac{8 \xi ^2}{\Delta _1^2 z_3^2-4 \xi ^2 \omega ^2}\right)}{3 m \xi } \text{tr}[{\bf M}_{12;12}]\\ \notag
 &+ \frac{\xi ^3 z_3^2 \left(4 m^2-t\right)  \left(\omega ^2 \left(4 m^2 \xi ^2 (\xi -\eta )+\eta  t\right)+m^2 \xi ^3 z_3^2 \left(4 m^2-t\right)\right)}{4 \Delta _1^2 \eta ^3 m t \omega ^4}\text{tr}[i \gamma _5{\bf M}_{01;02} + i \gamma _5{\bf M}_{02;01}]\\ \notag
 & + \frac{\xi  z_3 (\eta -\xi ) \left(t-4 m^2 \xi ^2\right) }{2 \Delta _1 \eta ^2 m t \omega }\text{tr}[i \gamma _5{\bf M}_{01;12}- i \gamma _5{\bf M}_{12;01}]  \\ \notag
 &  -\frac{\xi ^2  \left(z_3^2 \left(5 m^2-2 t\right)+5 \omega ^2\right) \left(4 \omega ^2 (\xi -\eta )+\xi  z_3^2 \left(4 m^2-t\right)\right)}{12 \Delta _1^2 \eta ^3 m \omega ^4} \text{tr}[i \sigma_{10}{\bf M}_{02;12}]\\ \notag
 &- \frac{1}{12 \Delta _1^2 \eta ^3 m \omega ^4 \big(4 \xi ^2 \omega ^2-\Delta _1^2 z_3^2\big)}\bigg(\xi ^2  \big(z_3^2 \big(5 m^2-2 t\big)+5 \omega ^2\big) \big(16 \xi ^2 \omega ^4 (\eta -\xi )+4 \omega ^2 z_3^2 \big(4 \eta  m^2 \xi ^2\\ \notag
 &+t \big(-\eta  \xi ^2+\eta +\xi \big)\big)+\xi  z_3^4 \big(4 m^2-t\big) \big(4 m^2 \xi ^2-\xi ^2 t+t\big)\big)\bigg)\text{tr}[i \sigma_{10}{\bf M}_{12;02}]\\ \notag
 & -\frac{\xi  t z_3^2 (\eta -\xi ) }{4 \Delta _1^2 \eta ^2 m \omega ^2}\text{tr}[i \sigma_{20}{\bf M}_{12;01}]\\ \notag
 &+ \frac{\xi ^2 t z_3^2 }{4 \Delta _1^2 \eta ^2 m \omega ^2}\text{tr}[i \sigma_{21}{\bf M}_{01;02}+ i \sigma_{21}{\bf M}_{02;01}]\\ \notag
 &-\frac{z_3^2  \left(z_3^2 \left(5 m^2-2 t\right)+5 \omega ^2\right) \left(4 \xi ^2 \omega ^2+\Delta _1^2 z_3^2\right)}{12 \eta ^2 m \omega ^4 \left(4 \xi ^2 \omega ^2-\Delta _1^2 z_3^2\right)}\text{tr}[i \sigma_{30}{\bf M}_{12;12}]\\ \notag
 &-\frac{5 (\eta -\xi ) }{3 \Delta _1^2 \eta  m} \text{tr}[i \sigma_{31}{\bf M}_{02;12}]\\ \notag
 &-\frac{5  \left(4 \xi ^2 \omega ^2 (\eta -\xi )+\Delta _1^2 z_3^2 (\eta +\xi )\right)}{3 \Delta _1^2 \eta  m \left(\Delta _1^2 z_3^2-4 \xi ^2 \omega ^2\right)} \text{tr}[i \sigma_{31}{\bf M}_{12;02}]\\ \notag
 &+\frac{\xi   \left(t z_3^5 \left(\xi ^2 \left(t-4 m^2\right)-t\right)+4 \xi ^2 t \omega ^2 z_3^3\right)}{8 \eta ^2 m \omega ^3 \left(4 \Delta _1 \xi ^2 \omega ^2-\Delta _1^3 z_3^2\right)}\text{tr}[i \sigma_{31}{\bf M}_{12;12}]\\ 
 &-\frac{\xi ^2 t z_3^2 (\eta -\xi ) }{4 \Delta _1^2 \eta ^3 m \omega ^2}\text{tr}[i \sigma_{32}{\bf M}_{01;12}]
\end{align}
%%%%%%%
%%%%%%%

\begin{align} \notag
{\rm Projection~onto}~&\wt{\frak{E}}_g:\\ \notag
 (\mathcal{P}[\wt{\frak{E}}_g],M) &=-\frac{4 m \xi  z_3  \left(4 \omega ^2 (\xi -\eta )+\xi  z_3^2 \left(4 m^2-t\right)\right)}{3 \Delta _1 \eta ^2 t \omega ^3}\text{tr}[{\bf M}_{02;12}]\\ \notag
 &-\frac{4 m \xi  z_3  \left(16 \xi ^2 \omega ^4 (\eta -\xi )+4 \omega ^2 z_3^2 \left(4 \eta  m^2 \xi ^2+t \left(-\eta  \xi ^2+\eta +\xi \right)\right)+\xi  z_3^4 \left(4 m^2-t\right) \left(4 m^2 \xi ^2-\xi ^2 t+t\right)\right)}{3 \Delta _1 \eta ^2 t \omega ^3 \left(4 \xi ^2 \omega ^2-\Delta _1^2 z_3^2\right)}\text{tr}[{\bf M}_{12;02}] \\ \notag
 &+\frac{8 m z_3^2  \left(\frac{8 \xi ^2}{4 \xi ^2 \omega ^2-\Delta _1^2 z_3^2}-\frac{1}{\omega ^2}\right)}{3 \xi  t} \text{tr}[{\bf M}_{12;12}]\\ \notag
 &+\frac{1}{4 \Delta _1^2 \eta ^3 t^2 \omega ^4} m \xi ^2 \bigg(-32 \xi  \omega ^4 (\eta -\xi ) \big(4 m^2 \xi ^2-\xi ^2 t+t\big)-4 \omega ^2 z_3^2 \big(32 m^4 \xi ^3 (\eta -2 \xi )+4 m^2 \xi  t \big(\eta  \big(4-5 \xi ^2\big)\\ \notag
 &+9 \xi ^3-6 \xi \big)+t^2 \big(\eta  \big(3 \xi ^2-4\big) \xi -5 \xi ^4+6 \xi ^2-1\big)\big)+z_3^4 \big(128 m^6 \xi ^4-16 m^4 \xi ^2 \big(7 \xi ^2-4\big) t\\ \notag 
 &+4 m^2 \big(8 \xi ^4-8 \xi ^2+1\big) t^2+\big(-3 \xi ^4+4 \xi ^2-1\big) t^3\big)\bigg) \text{tr}[i \gamma _5{\bf M}_{01;02}]\\ \notag
 &-\frac{m \xi ^2  \left(z_3^2 \left(8 m^2 \xi ^2+\left(2-3 \xi ^2\right) t\right)+8 \xi ^2 \omega ^2\right) \left(4 \omega ^2 (\xi -\eta )+\xi  z_3^2 \left(4 m^2-t\right)\right)}{2 \Delta _1 \eta ^3 t^2 \omega ^3 z_3} \text{tr}[i \gamma _5{\bf M}_{01;12}]\\ \notag
 &-\frac{m \xi ^2 z_3^2  \left(4 \omega ^2 \left(4 m^2 \xi ^3 (\xi -\eta )+t \left(\eta  \xi ^3-\xi ^4+1\right)\right)+z_3^2 \left(16 m^4 \xi ^4+m^2 \left(4-8 \xi ^4\right) t+\left(\xi ^4-1\right) t^2\right)\right)}{4 \Delta _1^2 \eta ^3 t \omega ^4}\text{tr}[i \gamma _5{\bf M}_{02;01}]\\ \notag
 &-\frac{4 \Delta _1 m z_3  \left(\frac{4 \omega ^2 (\eta -\xi )}{\xi }+z_3^2 \left(t-4 m^2\right)\right)}{t^2 \left(\omega  z_3^2 \left(t-4 m^2\right)-4 \omega ^3\right)}\text{tr}[i \gamma _5{\bf M}_{03;02}]\\ \notag
 &-\frac{8 m \xi  (\eta -\xi ) }{\eta  t^2} \text{tr}[i \gamma _5{\bf M}_{03;12}+ i \gamma _5{\bf M}_{13;02}]\\ \notag
 &+\frac{16 m \xi  \omega   \left(4 \omega ^2 (\xi -\eta )+\xi  z_3^2 \left(4 m^2-t\right)\right)}{\Delta _1 t^2 z_3 \left(z_3^2 \left(t-4 m^2\right)-4 \omega ^2\right)}\text{tr}[i \gamma _5{\bf M}_{13;12}]\\ \notag
 &+\frac{4 m \xi  (\eta -\xi )  \left(z_3^2 \left(5 m^2-2 t\right)+5 \omega ^2\right)}{3 \Delta _1^2 \eta ^2 t \omega ^2}\text{tr}[i \sigma_{10}{\bf M}_{02;12}]\\ \notag
 &-\frac{4 m \xi   \left(z_3^2 \left(5 m^2-2 t\right)+5 \omega ^2\right) \left(4 \xi ^2 \omega ^2 (\eta -\xi )+\Delta _1^2 z_3^2 (\eta +\xi )\right)}{3 \Delta _1^2 \eta ^2 t \omega ^2 \left(4 \xi ^2 \omega ^2-\Delta _1^2 z_3^2\right)}\text{tr}[i \sigma_{10}{\bf M}_{12;02}]\\ \notag
 &+\frac{m \xi ^2 z_3^2  \left(4 \omega ^2 (\xi -\eta )+\xi  z_3^2 \left(4 m^2-t\right)\right)}{4 \Delta _1^2 \eta ^3 \omega ^4}\text{tr}[i \sigma_{20}{\bf M}_{12;01}]\\ \notag
 &-\frac{m \xi ^2 z_3^2 }{\Delta _1^2 \eta ^2 \omega ^2}\text{tr}[i \sigma_{21}{\bf M}_{01;02}- i \sigma_{21}{\bf M}_{02;01}]\\ \notag
 &-\frac{m z_3^2  \left(z_3^2 \left(5 m^2-2 t\right)+5 \omega ^2\right) \left(4 \xi ^2 \omega ^2+\Delta _1^2 z_3^2\right)}{3 \eta ^2 t \omega ^4 \left(z_3^2 \left(\left(\xi ^2-1\right) t-4 m^2 \xi ^2\right)-4 \xi ^2 \omega ^2\right)}\text{tr}[i \sigma_{30}{\bf M}_{12;12}]\\ \notag
 &+\frac{20 m (\eta -\xi ) }{3 \Delta _1^2 \eta  t}\text{tr}[i \sigma_{31}{\bf M}_{02;12}] +\frac{20 m  \left(4 \xi ^2 \omega ^2 (\eta -\xi )+\Delta _1^2 z_3^2 (\eta +\xi )\right)}{3 \Delta _1^2 \eta  t \left(\Delta _1^2 z_3^2-4 \xi ^2 \omega ^2\right)}\text{tr}[i \sigma_{31}{\bf M}_{12;02}]\\ 
 &-\frac{ \left(\Delta _1^2 m \xi  z_3^5+4 m \xi ^3 \omega ^2 z_3^3\right)}{8 \Delta _1 \eta ^2 \xi ^2 \omega ^5-2 \Delta _1^3 \eta ^2 \omega ^3 z_3^2}\text{tr}[i \sigma_{31}{\bf M}_{12;12}]+\frac{m \xi ^2 z_3^2 (\eta -\xi ) }{\Delta _1^2 \eta ^3 \omega ^2}\text{tr}[i \sigma_{32}{\bf M}_{01;12}]
\end{align}
%%%%%%%
%%%%%%%

\begin{align}\notag
{\rm Projection~onto~} \frak{H}_g{}^T:\\ \notag
    \text{($\mathcal{P}$[}\frak{H}_g{}^T\text{],M)}&=-\frac{m \xi  z_3^2  \left(4 m^2 \xi ^2 (\eta -\xi )+t \left(-\eta  \xi ^2+\eta +\xi ^3\right)\right)}{\Delta _1^2 \eta ^2 t \omega ^2} \text{tr}[i \gamma _5{\bf M}_{01;02}+i \gamma _5{\bf M}_{02;01}]\\ 
    & +\frac{2 m \xi ^2 z_3 (\eta -\xi ) }{\Delta _1 \eta ^2 t \omega }\text{tr}[i \gamma _5{\bf M}_{01;12} + i \gamma _5{\bf M}_{12;01}]
\end{align}

%%%%%%%
%%%%%%%

Projection onto $\frak{E}_g{}^T$:
\begin{align}\notag
    \text{($\mathcal{P}$[}\frak{E}_g{}^T\text{],M)}&= \frac{8 m \xi ^2 z_3^2 \left(t-4 m^2\right) }{3 \Delta _1^2 \eta ^2 t \omega ^2} \text{tr}[{\bf M}_{01;01}-{\bf M}_{02;02}+{\bf M}_{12;12}]\\ \notag
    & +\frac{4 m^3 \xi  z_3^2 }{\Delta _1^2 \eta  t \omega ^2}\text{tr}[i \gamma _5{\bf M}_{01;02}+ i \gamma _5{\bf M}_{02;01}] \\ 
    &\frac{4 m \xi  z_3^2 \left(2 t-5 m^2\right) }{3 \Delta _1^2 \eta ^2 t \omega ^2} \text{tr}[i \sigma_{30}{\bf M}_{01;01} - i \sigma_{30}{\bf M}_{02;02} + i \sigma_{30}{\bf M}_{12;12}] 
\end{align}

%%%%%%%
%%%%%%%
Projection onto $\wt{\frak H}_g{}^T$:
\begin{align}
    \text{($\mathcal{P}$[}\wt{\frak H}_g{}^T\text{],M)}&=\frac{16 m^3 \xi ^2 z_3^2 }{3 \Delta _1^2 \eta ^2 t \omega ^2} \text{tr}[{\bf M}_{01;01}-{\bf M}_{02;02}+ {\bf M}_{12;12}] \\ \notag
    &-\frac{2 m^3 \xi  z_3^2 }{\Delta _1^2 \eta  t \omega ^2}\text{tr}[i \gamma _5{\bf M}_{01;02}+i \gamma _5{\bf M}_{02;01}] \\ 
    &+\frac{10 m^3 \xi  z_3^2 }{3 \Delta _1^2 \eta ^2 t \omega ^2}\text{tr}[i \sigma_{30}{\bf M}_{01;01}-i \sigma_{30}{\bf M}_{02;02}+i \sigma_{30}{\bf M}_{12;12}]
\end{align}

Projection onto $\wt{\frak E}_g{}^T$:
\begin{align}\notag
    \text{($\mathcal{P}$[}\wt{\frak E}_g{}^T\text{],M)}&=\frac{8 \xi ^3 z_3^2 \left(m t-4 m^3\right) }{3 \Delta _1^2 \eta ^2 t \omega ^2}\text{tr}[{\bf M}_{01;01}-{\bf M}_{02;02}] \\ \notag
    &+\frac{4 m \xi  z_3  \left(4 \omega ^2 (\xi -\eta )+\xi  z_3^2 \left(4 m^2-t\right)\right)}{3 \Delta _1 \eta ^2 t \omega ^3} \text{tr}[{\bf M}_{02;12}+{\bf M}_{12;02}] \\ \notag
    &+\frac{8 m z_3^2  \left(\Delta _1^2 \left(\eta  t z_3^2 (\eta -2 \xi )-4 \xi ^2 \omega ^2 (\eta -\xi )^2\right)-4 \eta ^2 \xi ^2 t \omega ^2\right)}{3 \Delta _1^2 \eta ^2 \xi  t \omega ^2 \left(t z_3^2-4 \xi ^2 \omega ^2\right)}\text{tr}[{\bf M}_{12;12}]\\ \notag
    &-\frac{4 m \xi  (\eta -\xi )  \left(z_3^2 \left(5 m^2-2 t\right)+5 \omega ^2\right)}{3 \Delta _1^2 \eta ^2 t \omega ^2}\text{tr}[i \sigma_{10}{\bf M}_{02;12}+i \sigma_{10}{\bf M}_{12;02}] \\ \notag
    &-\frac{m \xi ^3 z_3^2 (\eta -\xi ) \left(4 m^2-t\right) }{\Delta _1^2 \eta ^2 t \omega ^2}\text{tr}[i \sigma_{20}{\bf M}_{12;01}]\\ \notag
    &+ \frac{4 m z_3^2 }{t z_3^2 \left(t-4 m^2\right)-4 t \omega ^2}\text{tr}[i \sigma_{21}{\bf M}_{01;02}+i \sigma_{21}{\bf M}_{02;01}]\\ \notag
    &+\frac{2 m \xi ^2 z_3 (\eta -\xi ) }{\Delta _1 \eta ^2 t \omega }\text{tr}[i \sigma_{21}{\bf M}_{01;12}+i \sigma_{21}{\bf M}_{12;01}] \\ \notag
    & +\frac{4 \xi ^2 z_3^2 \left(2 m t-5 m^3\right) }{3 \Delta _1^2 \eta ^2 t \omega ^2}\text{tr}[i \sigma_{30}{\bf M}_{01;01}- i \sigma_{30}{\bf M}_{02;02}] \\ \notag
    &-\frac{1}{12 \Delta _1^2 \eta ^3 \xi ^2 t \omega ^4 \big(t z_3^2-4 \xi ^2 \omega ^2\big)}\big(m z_3^2  \big(\Delta _1^2 \big(80 \eta  \xi ^2 \omega ^4 (\eta -\xi )^2+3 \xi ^2 t^2 z_3^4 (\eta -2 \xi )\\ \notag
    &+4 t \omega ^2 z_3^2 \big(-5 \eta ^3+10 \eta ^2 \xi +6 \xi ^5\big)\big)+16 \eta ^3 \xi ^2 \big(3 \xi ^2+5\big) t \omega ^4\big)\big)\text{tr}[i \sigma_{30}{\bf M}_{12;12}] \\ \notag
    &-\frac{20 m (\eta -\xi ) }{3 \Delta _1^2 \eta  t} \text{tr}[i \sigma_{31}{\bf M}_{02;12}+i \sigma_{31}{\bf M}_{12;02}]\\ \notag
    &-\frac{m \xi  z_3^3 (\eta -2 \xi )}{2 \Delta _1 \eta ^3 \omega ^3} \text{tr}[i \sigma_{31}{\bf M}_{12;12}]+\frac{m \xi ^4 z_3^2 (\eta -\xi ) \left(4 m^2-t\right) }{\Delta _1^2 \eta ^3 t \omega ^2}\text{tr}[i \sigma_{32}{\bf M}_{01;12}]\\ 
    &+\frac{\xi ^4 z_3^3 \left(m t-4 m^3\right) }{2 \Delta _1 \eta ^3 t \omega ^3} \text{tr}[i \sigma_{32}{\bf M}_{01;02}+i \sigma_{32}{\bf M}_{02;01}]
\end{align}

\section{Forward case}
\label{app: forward}

It is also interesting to perform the treatment analogous to that used in this work, in the forward case. Indeed, we find that for generic hadron polarizations, there is a single linear relation between the possible Lorentz structures that can appear.

For $\Delta = 0$ and $z = z^- n$ the PDFs $f_g$ and $\tilde f_g$ are defined by
\begin{align}
M^{+i;+i} &= \frac{1}{2} f_g P^+ \bar u(p)\gamma^+ u(p),
\\
\widetilde M^{+i;+i} &= \frac{1}{2} \tilde f_g P^+ \bar u(p) i \gamma^+ \gamma_5 u(p).
\end{align}
We can take the same decomposition as in Eq.\,\eqref{eq: basis1}. Of course, the structures depending on $\Delta$ should be dropped. For the remaining 21 tensors we find a linear relation (leading to a basis of 20 tensors) which we can use to eliminate for example
\begin{align}
\mathcal T_{zz\sigma}^{\mu \nu; \alpha \beta} = i\bar u(p) ( z^{\alpha } z^{\mu } \sigma^{\beta \nu}  -z^{\beta } z^{\mu }  \sigma^{\alpha  \nu } -z^{\alpha } z^{\nu } \sigma ^{\beta  \mu }+z^{\beta } z^{\nu }
\sigma^{\alpha  \mu } )  u(p).
\end{align}
Then, we find that 
\begin{align}
f_g &= -2\mathcal M_{pp}  |_{z^2 = 0},
\\
\tilde f_g &= -2 \mathcal M_{pp \sigma} |_{z^2 = 0},
\end{align}
where the corresponding Lorentz structures are
\begin{align}
\mathcal T_{pp}^{\mu \nu; \alpha \beta} &= \bar u(p)  ( p^{\alpha } p^{\mu } g^{\beta  \nu }- p^{\beta } p^{\mu } g^{\alpha  \nu }- p^{\alpha } p^{\nu } g^{\beta  \mu }+ p^{\beta } p^{\nu } g^{\alpha  \mu } ) u(p),
\\
\mathcal T_{pp \sigma}^{\mu \nu; \alpha \beta} &= i\bar u(p)( p^{\beta } p^{\nu } \sigma^{\alpha  \mu }-p^{\alpha } p^{\nu }  \sigma^{\beta  \mu } -p^{\beta } p^{\mu } \sigma^{\alpha  \nu }+p^{\alpha } p^{\mu } \sigma^{\beta  \nu } )u(p).
\end{align}
Solving for the projections that obey
\begin{align}
(\mathcal P[f_g], M) = -2\mathcal M_{pp}, \qquad (\mathcal P[\tilde f_g], M) = -2 \mathcal M_{pp \sigma},
\end{align}
yields
\begin{align}
(\mathcal P[f_g],M) &= \frac{4}{5mp_0^2} \, \text{tr}\left [\textbf{M}_{02;02} + \textbf{M}_{12;12} \right ],
\\
(\mathcal P[\tilde f_g],M) &= \frac{1}{2m^3} \, \text{tr}\left [ i \sigma_{12} \left ( \textbf{M}_{02;01} + \textbf{M}_{23;13} + \frac{p_3}{p_0} \textbf{M}_{23;01} + \frac{p_3}{p_0}  \textbf{M}_{02;13}\right ) \right ].
\end{align}

%%%%%
%%%%%
%%%%%
%\clearpage
\bibliography{References.bib}
%\clearpage

%%%%%%%%%%%%%%%%%%%%%%%%%%
%%%%%%%%%%%%%%%%%%%%%%%%%%

\end{document}